\documentclass[journal,12pt,draftclsnofoot,onecolumn,twoside]{IEEEtran}
\normalsize

\usepackage{lscape}
\usepackage{color,soul}
\usepackage[normalem]{ulem} % either use this (simple) or

\usepackage{ifpdf}
 \ifpdf
     \usepackage[pdftex]{graphicx}
 \else
     \usepackage[dvips]{graphicx}
 \fi

\graphicspath{{figures/}}

\usepackage{cite}

\usepackage{epstopdf}					% convert eps figure automatically to pdf for pdflatex
 \usepackage{booktabs}

\usepackage[cmex10]{amsmath}
\usepackage{amssymb}

\usepackage{algorithm}

\usepackage{algorithm}
\usepackage{algpseudocode}

\usepackage{xcolor}
\usepackage{mdwmath}
\usepackage{mdwtab}
\usepackage{multirow}

\usepackage{caption}
\usepackage{subcaption}

\usepackage{tikz}
\usetikzlibrary{mindmap}
\usepackage{color}
\usepackage{enumitem}
\newlist{abbrv}{itemize}{1}
\setlist[abbrv,1]{label=,labelwidth=0.85in,align=parleft,itemsep=0.1\baselineskip,leftmargin=!}

\begin{document}
\title{Joint Inter-flow Network Coding and Opportunistic Routing in Multi-hop Wireless Mesh Networks:\\ A Comprehensive Survey}

\author{Somayeh~Kafaie,~\IEEEmembership{Student Member,~IEEE,} Yuanzhu~Chen,~\IEEEmembership{Member,~IEEE,} Octavia~A.~Dobre,~\IEEEmembership{Senior Member,~IEEE}, and Mohamed~Hossam~Ahmed,~\IEEEmembership{Senior Member,~IEEE}
    % <-this % stops a space

\thanks{This work was supported in part by the Natural Sciences and Engineering Research Council of Canada through its
Discovery program.}

\thanks{Somayeh Kafaie and Yuanzhu Chen are with the Department of Computer Science, Memorial University of Newfoundland,  St. John's, NL A1B 3X5, Canada. \protect\\E-mail: \{somayeh.kafaie, yzchen\}@mun.ca.}

\thanks{Octavia A. Dobre and Mohamed Hossam Ahmed are with the Faculty of Engineering and Applied Science, Memorial University of Newfoundland, St. John's, NL A1B 3X5, Canada. \protect\\
Email: \{odobre, mhahmed\}@mun.ca.}
}

\maketitle

\begin{abstract}

Network coding and opportunistic routing are two recognized innovative ideas to improve the performance of wireless networks by utilizing the broadcast nature of the wireless medium. 
%Looking at drawbacks and benefits of these two techniques suggests them as two complementing techniques with synergistic effect to further boost the capacity of the network.
In the last decade, there has been considerable research on how to synergize inter-flow network coding and opportunistic routing in a single joint protocol outperforming each in any scenario. This paper explains the motivation behind the integration of these two techniques, and highlights certain scenarios in which the joint approach may even degrade the performance, emphasizing the fact that their synergistic effect cannot be accomplished with a naive and perfunctory combination. This survey paper also provides a comprehensive taxonomy of the joint protocols in terms of their fundamental components and associated challenges, and compares existing joint protocols. We also present concluding remarks along with an outline of future research directions.

\end{abstract}

% Note that keywords are not normally used for peer review papers.

%%% list of abb

%\tableofcontents

\section{Introduction\label{sec:introduction}}

 \IEEEPARstart{W}{ireless} mesh network (WMN)~\cite{WMN-Akyildiz-Comm2005, WMN-Akyildiz-Wiley2009} is a type of wireless communication networks aiming to realize the dream of a seamlessly connected world. In mesh infrastructure, radio nodes are connected via wireless links creating a multi-hop wireless network, and nodes can talk to each other and pass data over long distances. This is realized by forming long paths consisting of smaller segments and handing off data over mulitple hops. This cooperative data delivery is the key idea of mesh networks to share connectivity across a large area with inexpensive wireless technologies.

Despite these advancements, users' expectations rise fast, and new applications require higher throughput and lower delay~\cite{5GTactileInternet-Simsek-JSAC2016}. In addition, the performance of wireless networks is significantly restricted by interference, and the unreliability of the wireless channel. Also, it is adversely affected by the contention among different data flows and devices in sharing bandwidth and other network resources. However, since the last decade two promising approaches of ``Opportunistic Routing'' and ``Network Coding'' are proved to improve the performance of wireless networks significantly by creatively utilizing the broadcast nature of the wireless medium.

\emph{Network coding} (NC), more specifically \emph{inter-flow network coding} (IXNC), is the process of forwarding more than one packet in each transmission. Doing so, it increases the ``effective'' capacity of the network~\cite{XCOR-Koutsonikolas-Mobicom2008} and improves the throughput. 
\emph{Opportunistic routing}\footnote{Also called ``opportunistic forwarding'' in some research~\cite{ORSurvey-Liu-CM2009, BEND-Zhang-CNJournal2010, CoAOR-Hu-GLOBECOM2013, Inter+opp-Mehmood-ComputNetw2013, diversitySurvey-Bruno-CompComm2010}.} (OR) also benefits from the broadcast nature of wireless networks via path diversity.
In OR, in contrast to traditional forwarding, there is no fixed route, and nodes do not forward a packet to a specified pre-selected next-hop. In fact, a node first broadcasts the packet, and then the next-hop is selected among all neighbors that have received the packet successfully. In addition, as explained in Section~\ref{subsec:OR}, OR can reduce the total number of transmissions by exploiting long but possibly low-quality links. Doing so, OR can largely increase the packet delivery probability and network throughput. 

Given that IXNC and OR are two promising techniques in wireless networks, the following research questions have risen in recent years. In which scenarios, each one of IXNC and OR performs better? How to boost the performance of wireless networks even more by combining these two great ideas in a single protocol? How to select a routing metric and a forwarder prioritizing mechanism for OR such that IXNC recognizes more coding opportunities in the network leading to a higher throughput? As discussed in Section~\ref{sec:review}, studies on this subject show that this combination, if realized carefully, can enable further improvement in the network performance. 

In the last decade, a significant amount of research has been conducted on applying the idea of IXNC and/or OR in wireless networks, and several review papers discuss the available studies as well as future research directions on OR~\cite{ORSurvey-Liu-CM2009, diversitySurvey-Bruno-CompComm2010, ORSurvey-Boukerche-ACMSurvey2014, ORSurvey-Chakchouk-COMST2015, ORsurveyUnderwater-Menon-ICETT2016} or NC~\cite{NCTheory1-Yeung-CIT2006, NCTheory2-Yeung-CIT2006, NCFund-Fragouli-NET2007, NCIntro-Ho-Cambridge2008, NCSurvey-Yeung-Springer2008, NCSurveyEnergy-Wang-ICICS2009, NCAlgorithms-Langberg-NetCod2009, NCSuerveyApps-Matsuda-IEICE2011, NCAwareSurvey-Iqbal-NetCompApp2011, NCTheorySyrvey-Bassoli-COMST2013, NCSurveyApps-Fragouli-NET2013, NCSurveyRelay-Mohammed-NCA2013, NCSurveyMultimedia-Magli-TM2013, CodingSurveyWSN-Ostovari-2014, SurveyNCCRN-Farooqi-NetCompApp2014, SurveyInterFlow-Xie-CompNetworks2015}. However, despite the increasing interest in combination of IXNC and OR, no study, to the best of our knowledge, illustrates the pros and cons of merging these two great ideas, and compares the proposed solutions highlighting their fundamental components and associated challenges.

\begin{figure}[ht]
\centering
\includegraphics[scale=0.8]{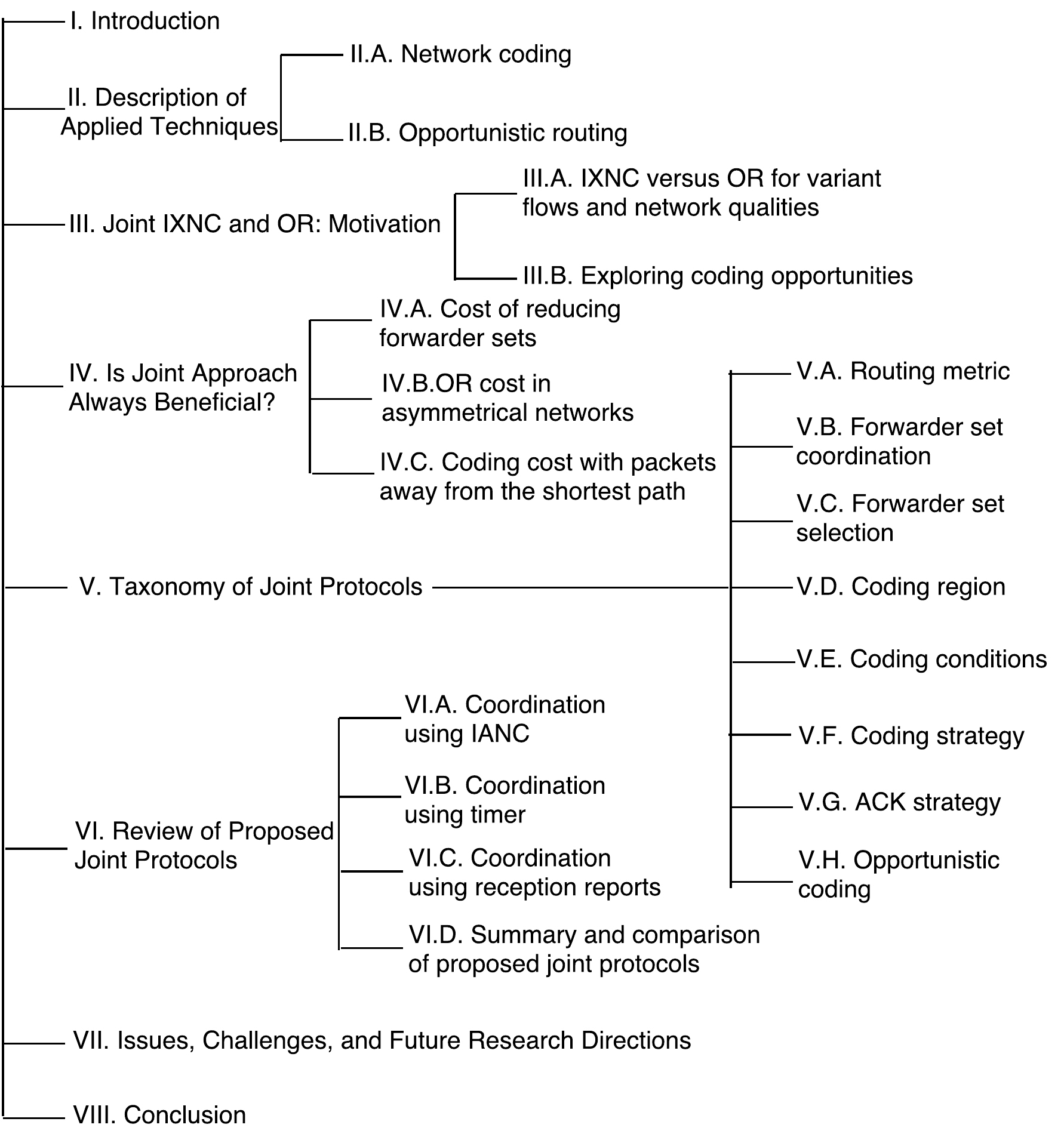}
\caption{The organization of the paper.}
\label{fig:outline}
\end{figure}

In this paper, we study the advantages and disadvantages of combining IXNC and OR especially in multi-hop wireless mesh networks, discuss the challenges and design criteria, and classify and compare the proposed joint IXNC and OR protocols.
The idea behind NC and OR is explained in Sections~\ref{sec:definiton}. The motivation for the integration of IXNC and OR and the benefits of the joint approach are illustrated in Section~\ref{sec:motivation}, while Section~\ref{sec:jointDrawbacks} discusses the cases that the joint approach may not outperform each of IXNC and OR individually. Section~\ref{sec:taxonomy} introduces the taxonomy of the joint protocols, in terms of the most important design decisions that should be made, and Section~\ref{sec:review} provides a brief overview of existing joint IXNC and OR protocols. Finally, Section~\ref{sec:future} presents open issues of the joint approach and promising future research directions, while Section~\ref{sec:conclusions} concludes the paper.
An overview of the overall organization of the paper is presented in \figurename~\ref{fig:outline}. 
Also, the terms used in this paper are summarized in Table~\ref{table:definition}, and the list of acronyms and abbreviations is provided at the end of the paper.

\begin{table}[!t]
\renewcommand{\arraystretch}{1.3}
\caption{Definition of some terms used in this article.}
\label{table:definition}
\centering
\begin{tabular}{|l|l|}
\hline
\textbf{Term} & \textbf{Definition} \\ \hline
\hline
native packet &  a packet not combined with any other packet \\  \hline
coded packet &  combination of more than one native packet \\  \hline
coding node &  a node in which coded packets are generated \\  \hline
coding partner &  a native packet encoded with other packets \\ \hline
\multirow{2}{*}{output queue} & a queue at each node storing the packets \\
& that should be forwarded \\ \hline
\multirow{2}{*}{reception report} & control packets sent by each node to advertise \\
 & its packet repository to its neighbors\\ \hline
\multirow{2}{*}{encoding vector}& a vector of coefficients representing the weight of \\
 &  the native packets in the IANC encoded packet\\ \hline
\multirow{2}{*}{forwarder set} & the set of next-hop candidates, which can be \\ 
 &  chosen as the next forwarder of a packet  \\ \hline
opportunistic & a case when the node overhears all  \\
 listening &  communications around over the wireless medium \\ \hline
\multirow{2}{*}{multi-hop coding} & a type of coding in which coded packets are not \\
 &  necessarily decoded at the next-hop \\ \hline
% decoding & the forwarder set of each coding partner excluding \\
% forwarder set & those nodes that cannot decode the packet\\ \hline
 decoding & the forwarder set of each coding partner in a \\
 forwarder set & coded packet excluding those nodes that cannot \\
 & decode the packet\\ \hline
 innovative & a packet which is linearly independent from the  \\
 packet & previously received packets \\ \hline
\end{tabular}
\end{table}

\section{Description of Applied Techniques \label{sec:definiton}}

\subsection{Network coding\label{subsec:NC}}
NC, introduced by Ahlswede \textit{et al.}~\cite{NC-Ahlswede-IEEETransactionsIT2000} in 2000, represents an effective idea to increase the transmission capacity of a data communication network as well as its robustness. 
In recent years, a significant amount of research has been conducted to explore the effect of NC in different scenarios and improve the network performance. 
A review of NC from the theoretical point of view can be found in~\cite{NCTheory1-Yeung-CIT2006, NCTheory2-Yeung-CIT2006, NCFund-Fragouli-NET2007, NCIntro-Ho-Cambridge2008, NCSurvey-Yeung-Springer2008, NCTheorySyrvey-Bassoli-COMST2013}. 
Matsuda \textit{et al.}~\cite{NCSuerveyApps-Matsuda-IEICE2011} discuss the NC design problem, its fundamental characteristics and its applications. 
Also, some studies review the applications of NC focusing on wireless and content distribution networks~\cite{NCSurveyApps-Fragouli-NET2013}, cognitive radio networks (CRN)~\cite{SurveyNCCRN-Farooqi-NetCompApp2014, CognitiveNCSurvey-Naeem-COMST2017} and wireless relay networks~\cite{NCSurveyRelay-Mohammed-NCA2013}.
A survey on unicast, multicast and broadcast applications of NC especially for wireless sensor networks (WSN) can be found in~\cite{CodingSurveyWSN-Ostovari-2014}, while Wang \textit{et al.}~\cite{NCSurveyEnergy-Wang-ICICS2009} discuss the benefits of NC in WSNs for energy saving. 
Furthermore, Langberg and Sprintson~\cite{NCAlgorithms-Langberg-NetCod2009} explain the proposed algorithms for NC and their computational complexity, and a review of the NC protocols and their associated issues from ``multimedia communication'' point of view can be found in~\cite{NCSurveyMultimedia-Magli-TM2013}.

In general, two different types of network coding can be applied, namely intra-flow network coding (IANC) and inter-flow network coding (IXNC)\footnote{Also called intra-session and inter-session~\cite{NCIntro-Ho-Cambridge2008, I2NC-Seferoglu-INFOCOM2011, NCSurveyMultimedia-Magli-TM2013}.}. In the context of this survey, a flow refers to a data stream (i.e., a sequence of data packets) between a given source and destination. Although both types of NC transmit coded packets in the network, IANC encodes packets of the same flow while IXNC applies encoding on packets of separate flows. Moreover, their goals and challenges are quite different as IANC is used to improve the robustness and reliability of wireless networks, while IXNC is utilized to boost the capacity of the network. 

\subsubsection{Intra-flow network coding \label{subsubsec:IANC}}
IANC increases the robustness by forwarding random linear combination of the packets of the same flow. It is an efficient alternative to the hop-by-hop feedback mechanism used in traditional forwarding in order to achieve reliability in the network. These feedback messages usually provide information about the packets that have already been received and the ones that should be retransmitted. By encoding packets originated from the same source, IANC makes all packets equally beneficial. Hence, it saves bandwidth by eliminating hop-by-hop feedback. 

In IANC, each node generates a linear combination of the packets of its transmission buffer over a finite field (e.g., $GF(2^{m})$), and sends the coded packet. Li \textit{et al.}~\cite{LNC-Li-IT2003} show that NC can still reach the broadcast capacity of the network even if the encoding of the packets is restricted to linear coding. In random linear network coding (RLNC), the node selects a vector of coefficients (i.e., encoding vector) representing the ``weight'' of the native packets in the encoded packet, and then combines the packets using addition and multiplication over the finite field. On the other hand, the destination keeps storing received encoded packets that provide new information (i.e., innovative packets) and their encoding vectors. They are stored as a matrix in the reduced row echelon form until the destination receives enough packets and can decode the packets using a technique like Gaussian elimination~\cite{GaussianElimination-Bareiss-MathComp68}. Then, the destination sends an acknowledgment (ACK) message notifying the source and intermediate nodes of receiving the packets.

\begin{figure}[ht]
\centering
\includegraphics[scale=0.9]{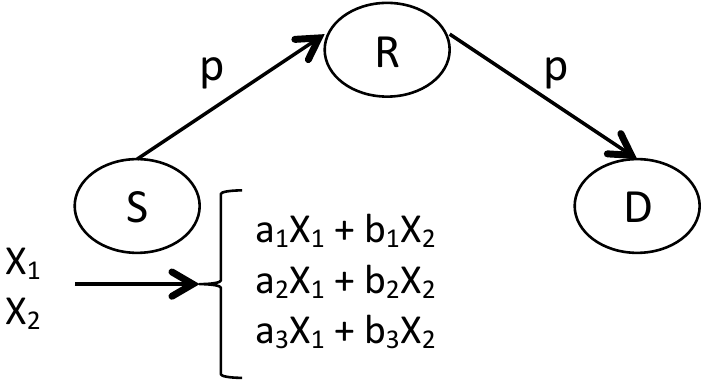}
\caption{IANC improves the reliability of the network by making all packets equally beneficial.}
\label{Intra-reliablity}
\end{figure}

To illustrate the idea, let us look at the example provided in \figurename~\ref{Intra-reliablity}, where the source $S$ needs to deliver two packets $X_{1}$ and $X_{2}$ to the destination $D$ via the relay $R$. Let us assume that the probability of successful transmission through each link is $p$. In a traditional scheme, after sending each packet to the relay if the source does not receive an ACK, it retransmits the same packet. However, in IANC instead of sending these two native packets separately, the source transmits $\lceil2/p\rceil$ coded packets in the form of $a_iX_{1}+b_iX_{2}, i \in \lbrace 1, 2, ..., \lceil2/p\rceil\rbrace, a_i,b_i \in GF(2^{m})$, consecutively, and then waits for an ACK. Similarly, $R$ generates linear combinations of the received packets and forwards them to the destination. As soon as the destination receives two innovative packets, it will be able to decode the native packets $X_{1}$ and $X_{2}$, and send back an ACK.

The idea of IANC has seen considerable attention from the research community and a significant amount of research has been conducted on IANC from both theoretical and practical point of views~\cite{BitTorrent-Gkantsidis-Infocom2005, ICEMAN-Joy-MobiCom2013, E-NCP-Lin-INFOCOM2008, MIXIT-Katti-SIGCOMM2008, CodeOR-Lin-ICNP2008, CCACK-Koutsonikolas-INFOCOM2010, ONCR-Xiang-INFOCOM2015}. 
MORE (MAC-independent Opportunistic Routing and Encoding)~\cite{MORE-Chachulski-SIGCOMM2007} is one of the first methods that realizes this idea in practical wireless scenarios. For more details on IANC and encoding and decoding using RLNC, we refer the reader to~\cite{Primer-Fragouli-SIGCOMM2006, NCGuidance-Kafaie-NECEC2013, Generation-Fragouli-Infocom2006, NC+DTN-Widmer-SIGCOMM2005, LNC-Li-IT2003}.

\subsubsection{Inter-flow network coding \label{subsubsec:IXNC}}
IXNC is a coding scheme in which a node combines the packets of multiple flows, using bitwise exclusive OR (\textit{XOR}) operation, and sends them at the same time through the channel. 
By reducing the number of transmissions, IXNC first improves the throughput and second decreases the interference in wireless channel. In fact, in IXNC by utilizing the broadcast nature of wireless networks, the network can reach its maximum capacity~\cite{COPE-Katti-IEEEACMTransactions2008}. 

\begin{figure}[ht]
\centering
\includegraphics[scale=0.9]{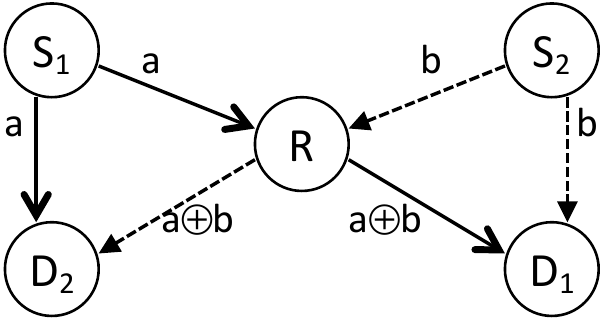}
\caption{X-topology showing how IXNC improves throughput.}
\label{fig:Xtopology}
\end{figure}

The practical research on IXNC in wireless networks for unicast communication was originally inspired by COPE~\cite{COPE-Katti-IEEEACMTransactions2008}. In COPE, every node is in promiscuous mode, and applies opportunistic listening (i.e., overhearing the transmission of other nodes). Also, each node sends the list of the packets that it has already received/overheard to its neighbors. These lists, called reception reports~\cite{COPE-Katti-IEEEACMTransactions2008}, are piggy backed on data packets or broadcast periodically. One of the most understood examples showing the gain behind IXNC is the X-topology in \figurename~\ref{fig:Xtopology}, where $S_{1}$ sends packet $a$ to $D_{1}$, and $S_{2}$ sends packet $b$ to $D_{2}$ through an intermediate node $R$. Since $D_1$ and $D_2$ are able to overhear the packets of the other flow from its source, the relay node $R$ mixes packets of two flows by applying \textit{XOR} and sends the coded packet $a \oplus b$ to the network. Then, for example, $D_2$ that has already overheard $a$ can decode $b$ by \textit{XOR}ing the received packets $a$ and $a \oplus b$. Doing so, IXNC decreases the number of required transmissions to deliver packets to their final destinations and improves the performance. Xie \textit{et al.}~\cite{SurveyInterFlow-Xie-CompNetworks2015} provide a survey on IXNC with unicast traffic under both reliable links and lossy links of WMNs, and discuss some drawbacks of IXNC in WMNs including spatial reuse reduction and lower link rate selection.

\subsection{Opportunistic routing\label{subsec:OR}}
OR~\cite{EXOR-Biswas-SIGCOMM2005, OPRAH-Westphal-MASS2006, GeRaF-Zorzi-TMC2003, MORE-Chachulski-SIGCOMM2007} is an effective idea to improve the performance of wireless networks, especially in lossy operation conditions, by providing more chances for a packet to make progress toward the destination (i.e., getting closer to the destination after the transmission in the sense of forwarding cost). In contrast to traditional forwarding in which the packets are forwarded along a fixed path, OR picks the next-hop of each packet only after the packet has been forwarded.

The idea behind OR, mostly recognized by ExOR (Extremely Opportunistic Routing)~\cite{EXOR-Biswas-SIGCOMM2005}, is that the route which packets traverse is not predetermined and can be different for each packet of the flow. In fact, the source selects a set of nodes which are closer to the destination than itself (i.e., called forwarder set), and after each packet transmission, the highest priority node that receives the packet will forward it toward the destination.
As discussed in Section~\ref{subsec:metric}, the nodes in the forwarder set are ranked based on a metric such as geographic distance, hop-count or ETX (Expected Transmission Count)~\cite{ETX-Couto-MobiCom2003}, which represents the expected number of transmissions for a packet.
Since, in OR, multiple nodes are candidates of receiving and forwarding a packet, one needs to coordinate the intermediate nodes and prevent forwarding multiple copies of the same packet (details in Section~\ref{subsec:coordination}). Therefore, the most important steps in OR are routing metric selection, forwarder set determination, and forwarder set coordination~\cite{ORSurvey-Boukerche-ACMSurvey2014}. 

Liu~\textit{et al.}~\cite{ORSurvey-Liu-CM2009} explain the technical challenges of the mentioned steps, and provide a survey of proposed OR protocols, while briefly discussing the joint OR and IXNC approach as ``coding-aware opportunistic routing mechanism''.
A survey of OR protocols for underwater WSNs can be found in~\cite{ORsurveyUnderwater-Menon-ICETT2016}. 
Also, Boukerche and Dareshoorzadeh~\cite{ORSurvey-Boukerche-ACMSurvey2014} review the issues of OR focusing 
on the mentioned three steps, and discuss the most important proposed OR protocols and analytical models together with the applications of OR in multicast and mobile scenarios.
Chakchouk~\cite{ORSurvey-Chakchouk-COMST2015} reviews the main components of OR, and provides a taxonomy of the proposed OR protocols in five different classes, namely the geographic, link-state-aware, probabilistic, optimization-based and cross-layer OR classes.
In another review paper, Bruno and Nurchis~\cite{diversitySurvey-Bruno-CompComm2010} not only discuss OR and network coding-aware (NC-aware) routing protocols but also the hybrid routing (i.e., the combination of OR and NC). However, their focus is mostly on the combination of IANC and OR, and just briefly mention merging IXNC with OR as ``neighborhood-based hybrid scheme''.

\begin{figure}[ht]
\centering
\includegraphics[scale=0.9]{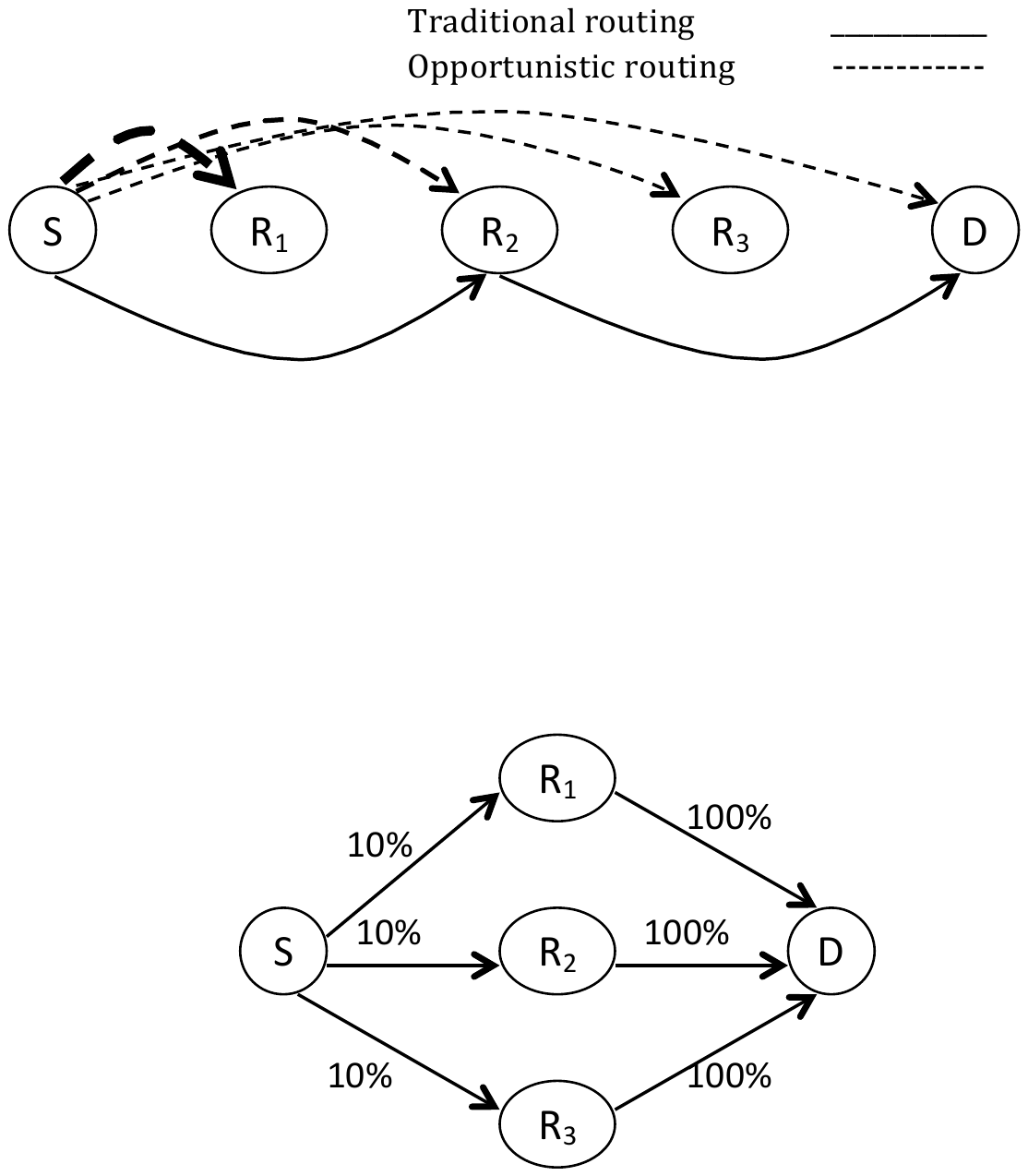}
\caption{Opportunistic routing and variant progress opportunities toward the destination.}
\label{fig:OR1}
\end{figure}

We use two examples, inspired by~\cite{EXOR-Biswas-SIGCOMM2005}, to elaborate on how OR improves the network performance. First, in \figurename~\ref{fig:OR1},
let us assume that node $S$ wants to deliver its packets to node $D$, 
and the link quality decreases with the distance between nodes. Any of the nodes, $R_{1}$, $R_2$, $R_{3}$ and $D$, may receive the packet sent by $S$ but with different probabilities depending on the quality of the link between each of them and the source. Hence, given that $P_{i,j}$ denotes the probability of successful transmission from node $i$ to node $j$, we have $P_{S,R_1}>P_{S,R_2}>P_{S,R_3}>P_{S,D}$.
In general, if $S$ chooses a node close to itself, like $R_1$, as the next-hop, the probability that the packet can be received is higher (i.e., the number of required retransmissions for $S$ to deliver a packet to $R_1$ is small), but the progress toward the destination is small (i.e., the packet is still far away from $D$). On the other hand, if $S$ selects a node close to the destination, like $R_3$, as the next-hop, the progress toward the destination is larger (i.e., the packet gets very close to $D$). However, the number of required retransmissions for $S$ to transmit a packet successfully to $R_3$ could be greater as well.
In traditional forwarding, $S$ selects a node with a good enough link, let it be $R_2$, as the next-hop. If the packet cannot be received by $R_2$, $S$ will retransmit the packet, even if it could have made smaller progress by reaching $R_1$, or was lucky enough to reach $R_3$ or even $D$. OR takes advantage of those smaller progresses as well as the lucky long strides by selecting all these nodes, $D$, $R_{3}$, $R_{2}$ and $R_{1}$, in the forwarder set, and as long as one of them receives the packet, $S$ does not need to retransmit it. Then, these nodes are coordinated, and the highest priority node (i.e., the closest node to the destination) that has received the packet will forward it.

\begin{figure}[ht]
\centering
\includegraphics[scale=0.9]{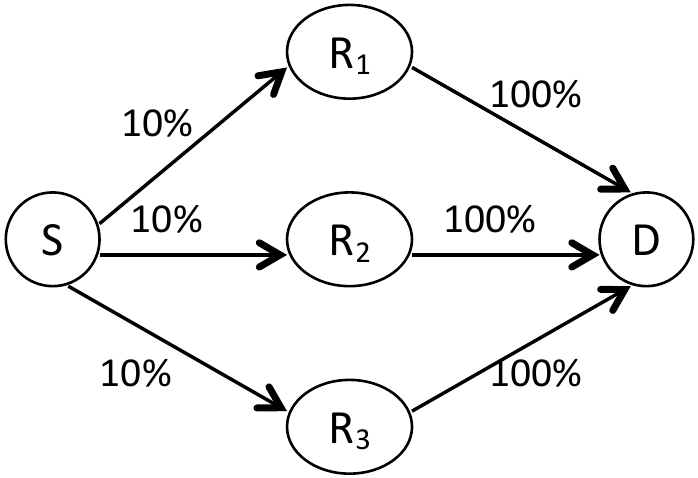}
\caption{Opportunistic routing and independent transmission opportunities.}
\label{fig:OR2}
\end{figure}

As another example, in \figurename~\ref{fig:OR2}, $S$ transmits packets to $D$ via a relay node. Let us assume that the link quality between the source and relays is poor, and only $10\%$ of the transmissions are successful, while relays can deliver all their packets successfully to the destination. In traditional forwarding, $S$ selects one of the nodes, $R_1$, $R_2$ or $R_3$ as the relay node, and the average number of required transmissions to deliver a packet to the destination is $1/0.1+1=11$ (i.e., an average of $10$ transmissions for each packet to be received by the relay and one to be delivered to the destination). On the other hand, in OR, as long as one of the intermediate nodes receives the packet, $S$ does not need to retransmit the packet. Given a perfect coordination among the relays, this reduces the number of required transmissions to deliver a packet to the destination to $\dfrac{1}{1-(1-0.1)^3}+1=4.69$, where $(1-(1-0.1)^3)$ represents the probability that at least one of the three intermediate nodes receives the packet. By providing such a strong virtual link (as combination of the real links)~\cite{ORSurvey-Hsu-CompNet2011}, OR can provide more independent chances for a packet to be received~\cite{EXOR-Biswas-SIGCOMM2005}, and reduce the number of required transmissions for packet delivery.

\section{Joint IXNC and OR: Motivation \label{sec:motivation}}
As explained before, both IXNC and OR improve the performance of wireless networks by utilizing the broadcast nature of the wireless medium. However, they 
have different applications, address separate challenges and may present even contrastive behaviors, especially regarding the effect of the number of flows on their performance, and the quality of the network for which these techniques are suitable. 
The goal of the joint approach is to merge their applications, and use the strengths of one to overcome the weakness of the other. For example, a major challenge of IXNC is finding an efficient way to explore coding opportunities in the network for which OR seems a promising solution.

\subsection{IXNC versus OR for variant flows and network qualities}

OR spreads packets across several nodes, and the packets might not meet at the same node to be combined, especially in lossy networks. That is to say, each node may receive a subset of the packets, and the packets that can be coded together might not be received by the same node. This means less traffic and perhaps less coding opportunities at IXNC for each node~\cite{O3-Han-MobiHoc2011}. Furthermore, in OR protocols like MORE~\cite{MORE-Chachulski-SIGCOMM2007} as the number of flows increases, the throughput gain of the protocol decreases~\cite{XCOR-Koutsonikolas-Mobicom2008, CAOR-Chung-ICC2012}, while in IXNC, more flows crossing at the same node provide more coding opportunities. 

IXNC is mostly effective in reliable networks with high-quality links, where nodes can rely on packet overhearing to decode received coded packets~\cite{XCOR-Koutsonikolas-Mobicom2008, Inter+opp-Mehmood-ComputNetw2013, INCOR:Inter+opp-Zhu-ICC2015}. 
It is not usually applicable in lossy environments since the accuracy of the coding node's estimaton of next hops' decoding ability decreases as the loss rate in the network increases. Due to this problem, COPE turns off IXNC if the loss rate in the network is higher than a threshold (i.e., the default value is 20\% in their implementation)~\cite{COPE-Katti-IEEEACMTransactions2008}. 

To illustrate the issue, let us assume in \figurename~\ref{fig:Xtopology}, $D_{2}$ cannot overhear a considerable number of sent packets from $S_{1}$ due to the loss of the link between $S_{1}$ and $D_{2}$. As we explained earlier, $R$ encodes the received packets of $S_{1}$ and $S_{2}$ together. However, because of the loss of overhearing link, $D_{2}$ cannot decode some received coded packets. In fact, $D_{2}$ cannot decode some packets like $b$ because it was not able to overhear corresponding packet $a$. In addition, with poor channel quality the reception reports can be lost easily, which makes encoding decisions more difficult. 

Although a few studies have been conducted to make IXNC efficient in lossy environments, they are not usually practical due to their computational complexity~\cite{CLONE-Rayanchu-SIGMETRICS2008, I2NC-Seferoglu-INFOCOM2011}, or they consider a scenario where every node transmits its packets to all other nodes~\cite{LossyNCPersistence-Munaretto-Symposium2007}.
Furthermore, some studies add IANC to IXNC~\cite{I2MIX-Qin-SECON2008, CandM-Zhu-IAS2009, Inter+Intra-Wang-Asilomar2009, I2NC-Seferoglu-INFOCOM2011, CORE-Hansen-CAMAD2013, CORE-Krigslund-VTC2013, polynomialInter+Intra-Khreishah-MASS2011,polynomialInter+Intra-Khreishah-ieeeTransPDS2013, MUFEC-Wang-PE2011, Rate+Partitioning+I2NC-Ou-Globecom2012} to improve its reliability and robustness in lossy networks. However, since these methods do not consider OR, they can only catch coding opportunities within the shortest path.

In contrast, OR scatters the packets of a flow over multiple paths from the source to the destination. In fact, by selecting more than one next-hop, OR provides more chances for a packet to make progress toward the destination, and can largely reduce the number of required transmissions and increase throughput, especially in lossy environments. Therefore, OR is mostly suitable for lossy environments with medium- to low-quality links between nodes, where restriction to a single next-hop may lead to packet losses and retransmissions~\cite{XCOR-Koutsonikolas-Mobicom2008, Inter+opp-Mehmood-ComputNetw2013, CAOR-Chung-ICC2012, INCOR:Inter+opp-Zhu-ICC2015}. 

\subsection{Exploring coding opportunities \label{subsec:joint}}
\begin{figure}[ht]
\centering
\includegraphics[scale=0.9]{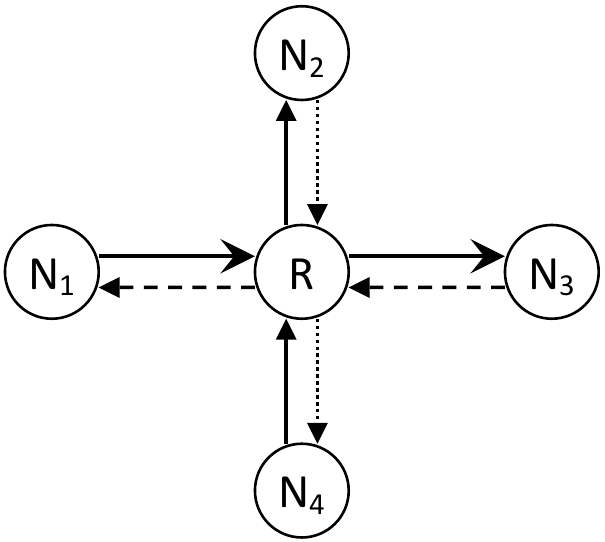}
\caption{Cross topology with 4 flows intersecting at $R$~\cite{COPE-Katti-IEEEACMTransactions2008}.}
\label{fig:cross}
\end{figure}

Coding opportunities in COPE, as one of the prominent examples of IXNC, are restricted only to joint nodes that receive packets from multiple flows. For example, let us assume that in the cross topology depicted in \figurename~\ref{fig:cross}, for each node all other nodes are in its transmission range except for the diametrically opposite one, and that $N_1$, $N_2$, $N_3$ and $N_4$ are the sources of 4 flows, intersecting at $R$, to the destinations $N_3$, $N_4$, $N_1$ and $N_2$, respectively. Then, $R$ can mix 4 packets received from all sources because each next-hop contains all other coding partners except for its intended packet. However, if the sources choose a different intermediate node than $R$, all flows cannot intersect at the same node and fewer coding opportunities are provided by COPE. 

This example shows that the improvement of throughput in protocols like COPE depends on the traffic pattern, 
which is totally independent from the potential coding opportunities. The flows, in IXNC, usually travel through the shortest path, and nodes passively wait for straightforward coding opportunities available in the designated routes without any ability to exploit potential coding opportunities beyond that~\cite{ANCHOR-Jiao-WiCom2008}. As a matter of fact, regular IXNC methods, like COPE, limit coding opportunities because their coding is passive and can be performed only at joint nodes.

To deal with the mentioned issue, NC-aware routing protocols~\cite{NCaware-Sengupta-Infocom2007, NCaware-Ni-WiMesh2006, NCaware-Sengupta-TN2010, DCAR-Le-TMC2010, FORMpostpone-Guo-VT2011, NCaware-Najjar-TWC2011, NCaware-Peng-GlobeCom2010, OCAR-Sun-China2010} are proposed. In these protocols, the route, through which packets travel, is selected with the awareness of available coding opportunities in the network, and the packets are forwarded via relays with more encoding chance. Therefore, network-coding-aware routing is similar to traditional routing in the sense that it chooses one single (best) route to forward all packets of a flow. However, as part of the routing metric to find the next-hop, it takes into account the number of available coding opportunities at each node.
Details on NC-aware routing protocols can be found in~\cite{NCAwareSurvey-Iqbal-NetCompApp2011}. 

In general, they are either centralized~\cite{NCaware-Sengupta-Infocom2007, NCaware-Ni-WiMesh2006, NCaware-Sengupta-TN2010} with scalability problem making their implementation in WMNs infeasible~\cite{CAR-Liu-NetSysManage2015, CORE-OR-Yan-IEEEWC2010}, or distributed~\cite{DCAR-Le-TMC2010, NCaware-Peng-GlobeCom2010, FORMpostpone-Guo-VT2011} with easier implementation in practical networks.
However, decisions in such NC-aware routing protocols are deterministic~\cite{HCOR-Hai-EURASIP2014}, and their chosen paths might be sub-optimal~\cite{CAR-Liu-NetSysManage2015} due to the traffic pattern and lossy characterisitics of the wireless network. For example, let us assume that at the beginning the best path for flow $A$ is $P_A$, where it can be mixed with flow $B$. However, when a new flow $C$ starts or flow $B$ ends, $P_A$ might not be the best path for flow $A$ anymore. If $A$ keeps using the same route, the network throughput drops, while searching for a better route and rerouting causes a long delay and degrades the network performance~\cite{HCOR-Hai-EURASIP2014}.

\begin{figure}
\centering
\begin{subfigure}{.35\textwidth}
  \centering
  \includegraphics[width=0.88\linewidth]{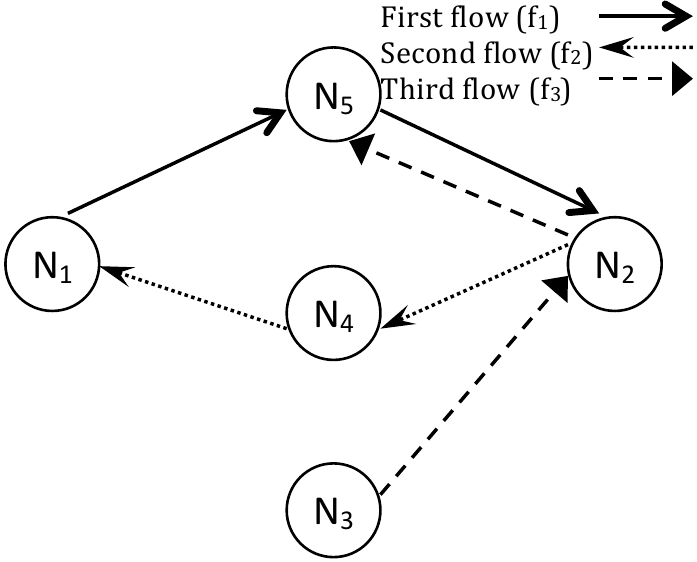}
  \caption{Traditional routing.}
  \label{fig:benefitsTraditional}
\end{subfigure}%
\begin{subfigure}{.35\textwidth}
  \centering
  \includegraphics[width=0.95\linewidth]{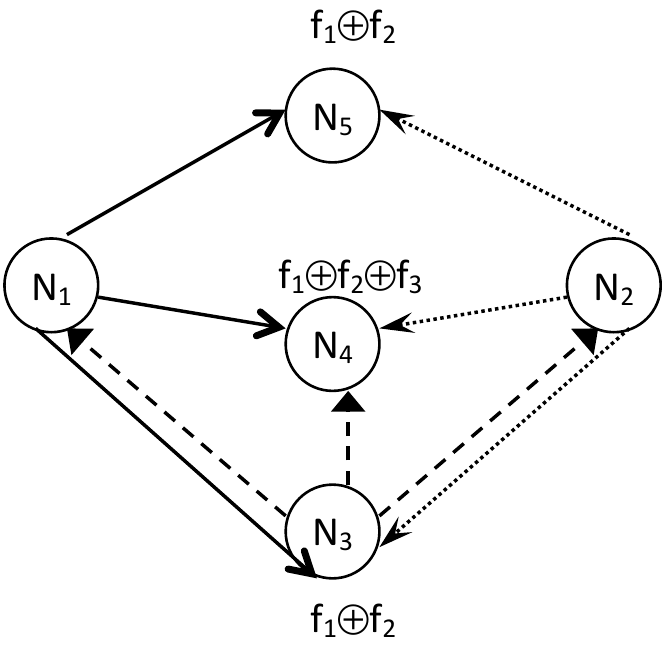}
  \caption{Joint OR and IXNC.}
  \label{fig:benefitsJoint}
\end{subfigure}
\caption{OR can improve the performance of IXNC.}
\label{fig:jointBenefit}
\end{figure}

To cope with the dynamics of the traffic and the network, NC-aware routing needs to be integrated with OR. To illustrate the idea, let us look at the example provided in \figurename~\ref{fig:jointBenefit}, reproduced with some changes from~\cite{CAR-Liu-NetSysManage2015}. At the beginning, there exist two flows in the network between $N_1$ and $N_2$, and the third flow from $N_3$ to $N_5$ is added later at time $t_1$. All nodes can overhear each other except for the mentioned endpoints (i.e., no direct link between $N_1$ and $N_2$, and also between $N_3$ and $N_5$). 
Because of the routing based on the shortest path, regular IXNC protocols, like COPE, cannot find any coding opportunity in this scenario, and their performance is not better than the traditional routing. Also, NC-aware routing may not provide the best performance especially if it selects $N_5$ or $N_3$ as the relay at the beginning, which in this case after joining $f_3$, it either should pay the cost of rerouting or keep routing on an inferior path. On the other hand, when IXNC is integrated with OR, since there is no designated route anymore, the performance can be improved by asking the node with the most coding opportunities to forward the packet in each transmission. For example, before $t_1$, each of the nodes $N_3$, $N_4$ or $N_5$ that overhears the packets of both flows is selected as the relay, and after $t_1$ if $N_5$ overhears all three flows, the joint approach chooses it correctly as the best forwarder.

Looking at pros and cons of OR and IXNC, one can think of them as two complementing techniques. 
OR can improve the performance of IXNC in lossy networks. 
In addition, in OR, the packet is first broadcast and then the next-hop is decided based on a metric. This metric, which prioritizes the possible forwarders of a packet, can be chosen so that the forwarder with more coding opportunities is selected in each transmission. Doing so and by considering the packets of the neighborhood collectively, OR can provide more coding opportunities in the network without forcing flows to cross at some focal nodes and causing faster energy drainage, longer delay, buffer overflow, and channel contention. In addition, by providing ``free-ride'' to packets of several flows, IXNC can improve the performance of OR in the presence of multiple flows. 

Therefore, there is an interesting research question on how to combine OR and IXNC such that their integration outperforms each of them individually in different scenarios. To realize such a powerful joint approach, the following challenges should be addressed:
\begin{itemize}
\item Choosing an appropriate routing metric to determine the set of forwarders, considering the specifications of both OR and IXNC.
\item Recognizing coding opportunities and selecting right packets to be coded together.
\item Prioritizing the candidates in the forwarder set and selecting the best one.
\item Coordinating the forwarder set and suppressing duplicate packets in the network. 
\end{itemize}

\section{Is the Joint Approach Always Beneficial? \label{sec:jointDrawbacks}}
In Section~\ref{sec:motivation}, we discussed the benefits of integrating the IXNC and OR techniques and explained the synergy between them. However, there are still scenarios in which the joint approach may not outperform each of these two techniques individually, or even degrades the performance especially if the forwarder set and/or the routing metric are not selected appropriately.

\subsection{Cost of reducing forwarder sets \label{subsec:shrinkage}}
Hai \textit{et al.}~\cite{HCOR-Hai-EURASIP2014} argue that the joint approach may perform worse than OR because of the shrinkage of the forwarder set. They define the cost of delivering packet $p$ from node $i$ to the destination via forwarder set $J$ as $C^p_i = c^p_{iJ} + C^p_J$. $c^p_{iJ}$ represents the cost of forwarding packet $p$ from $i$ to the forwarder set $J$ and is calculated as
\begin{equation}
c^p_{iJ}=\dfrac{1}{1-\prod_{j \in J}(1 - d_{ij})} \nonumber ,
\end{equation}
where $d_{ij}$ denotes the packet delivery ratio from node $i$ to node $j$. Intuitively, this cost (i.e., $c^p_{iJ}$) is inversely proportional to the probability that at least one of the nodes in the forwarder set receives the packet. Also, $C^p_J$ presents the weighted average of the delivery cost of packet $p$ from the nodes in the forwarder set $J$ to the destination.
They show that the cost of a forwarder set is optimal if and only if all nodes in the neighborhood with cost lower than the sender are included in the set. Intuitively, it means that if all neighbors of the sender, closer to the destination than the sender itself, are included in the forwarder set of a packet, the chance that the packet can make some progress toward the destination is maximum. Hence, not including some of such nodes or adding other nodes, which are farther from the destination than the sender, will reduce the chance. 

Given that, if $c^a_{iJ} + c^b_{iK}$ denotes the cost of forwarding packets $a$ and $b$ separately from node $i$ to their forwarder sets $J$ and $K$, respectively, the cost of forwarding them as a coded packet ($a \oplus b$) to the same forwarder sets equals 
\begin{equation}
c^a_{iJ} + c^b_{iK}-\dfrac{c^a_{iJ} \times c^b_{iK}}{c^a_{iJ} + c^b_{iK} -1} \nonumber .
\end{equation} 
Based on this equation, 
when a node forwards a coded packet like $a \oplus b$, the forwarding cost decreases because of the coding-gain and the ``free-ride'' transmission provided by IXNC. 
On the other hand, the forwarding cost might increase due to the shrinkage of the forwarder set. In fact, the forwarder set of each of packets $a$ and $b$, in the coded case (i.e., called decoding forwarder set) is a subset of their original forwarder set, when they were to be sent natively. The reason is that some of the nodes in the original forwarder set might not be able to decode the coded packet $a \oplus b$; thus not included in the forwarder set. 

\begin{figure}[ht]
\centering
\includegraphics[scale=0.9]{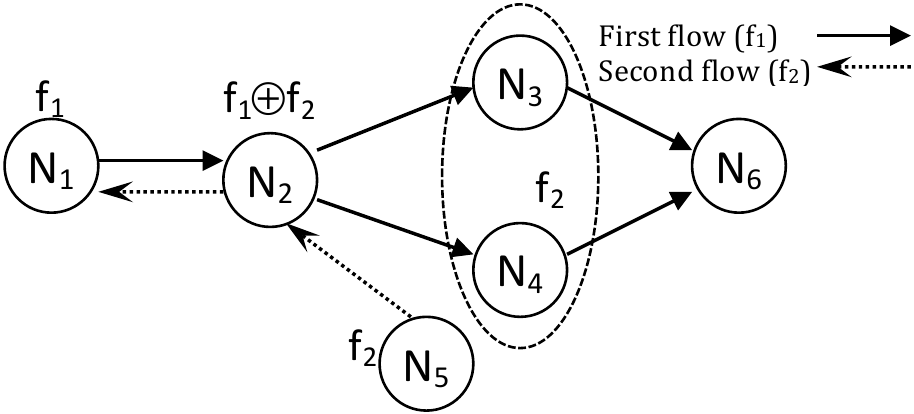}
\caption{Decoding forwarder set - how network coding may shrink the forwarder set of coding partners.}
\label{fig:decodingFS}
\end{figure}

\figurename~\ref{fig:decodingFS} shows how network coding may shrink the forwarder set of coding partners. In the scenario depicted in this figure, there exist two flows from $N_1$ and $N_5$ to $N_6$ and $N_1$, respectively. The figure shows a snapshot of the network representing the available packets at each node, where $N_4$ can overhear the packets of the second flow (i.e., $f_2$ from $N_5$ to $N_1$) from $N_5$. Thus, $N_2$ can encode the packets of the two flows. Originally, if $f_1$ was to be forwarded natively, $N_2$ could have chosen $N_3$ and $N_4$ in its forwarder set. However, when $f_1$ and $f_2$ are sent as a coded packet, only $N_4$ can decode $f_1$ and can be selected in the new forwarder set (i.e., decoding forwarder set).
As stated before, if any neighbor of the sender, closer to the destination than the sender itself, can not be included in the decoding forwarder set because of the decodability issue, the cost of the forwarder set will increase.

Thus, it is clear that in the joint approach, naively mixing packets does not necessarily improve the network performance, and the trade-off between coding gain and shrinkage cost should be resolved. In fact, forwarding coded packets is beneficial as long as its gain is greater than the price of forwarding the packets to the smaller decoding forwarder set.

\subsection{OR cost in asymmetric networks \label{subsec:asymmetric}}
Mehmood \textit{et al.}~\cite{Inter+opp-Mehmood-ComputNetw2013} highlight some scenarios in which IXNC over the shortest path may outperform IXNC integrated with OR. They study bidirectional unicasts flows when endpoints are out of transmission range of each other but they have some common neighbors relaying packets for them. A generalized scenario would be the case that intermediate nodes are partitioned in disjoint groups, where the nodes in each group are the same number of hops away from each endpoint. Each node can only communicate directly with the nodes in its group or adjacent groups, and no node includes nodes from its own group in the forwarder set even if they are closer to the destination (e.g., in terms of ETX or geo-distance) than the node itself. In such a network, they argue that naive joint methods are sub-optimal unless all intermediate nodes have identical link quality to both endpoints (i.e., symmetric network).

\begin{figure}[ht]
\centering
\includegraphics[scale=0.9]{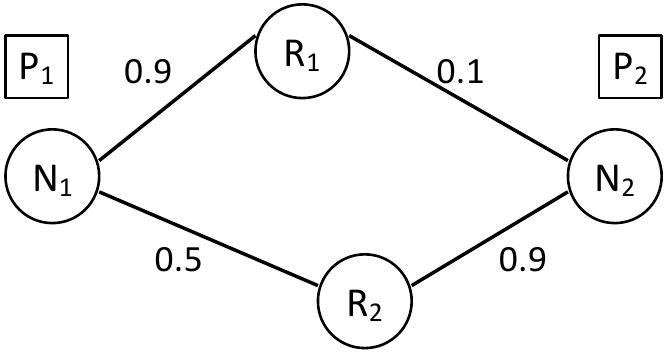}
\caption{Joint approach is not always the best one.}
\label{fig:IXNCvsJoint2}
\end{figure}

The scenario depicted in \figurename~\ref{fig:IXNCvsJoint2}, reproduced from~\cite{Inter+opp-Mehmood-ComputNetw2013} with slight changes, describes a case in which IXNC over a pre-chosen relay outperforms the joint approach. Let us assume that $N_1$ and $N_2$ send packets $P_1$ and $P_2$ to each other via a relay ($R_1$ or $R_2$), and the relays cannot overhear each other or do not consider each other in their forwarder set. Also, the label of each link represents the probability of successful transmission over that link.

Applying regular IXNC, it is clear that $R_2$ is the best relay, and $N_1$ and $N_2$ exchange their packets by $\dfrac{2}{0.9}+\dfrac{2}{0.5}-\dfrac{1}{1-(1-0.9)(1-0.5)} \approx 5.17$ expected number of transmissions (ExNT). The first two terms represent the ExNT to successfully forward the packets from the sources to $R_2$, and from $R_2$ to the destinations, respectively (without applying IXNC), and the last term denotes the coding gain. 

On the other hand, in the joint approach, it is likely that $R_1$ receives $P_1$ from $N_1$, and $R_2$ receives $P_2$ from $N_2$, and without any coding opportunity they send the packets separately. Given the poor quality link between $R_1$ and $N_2$, about $8.857$ ExNTs are required in this case to deliver both packets to their destinations. We refer the reader for calculation details to \cite{Inter+opp-Mehmood-ComputNetw2013}. Comparing the expected number of required transmissions for both IXNC and the joint approach shows that regular IXNC over the shortest path may outperform the naive joint approach in such an asymmetric network.
%The details have been provided in~\cite{Inter+opp-Mehmood-ComputNetw2013}. 

\subsection{Coding cost with packets far away from the shortest path \label{subsec:shortestPath}}
\begin{figure}
\centering
\begin{subfigure}{.35\textwidth}
  \centering
  \includegraphics[width=0.95\linewidth]{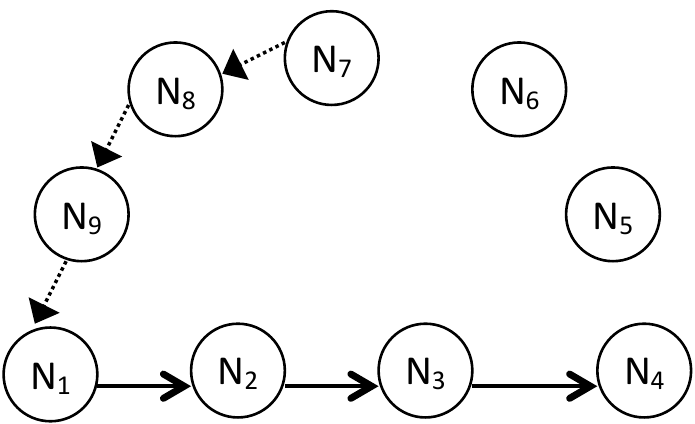}
  \caption{Traditional routing.}
  \label{fig:notBeneficialTraditional}
\end{subfigure}%
\begin{subfigure}{.35\textwidth}
  \centering
  \includegraphics[width=0.95\linewidth]{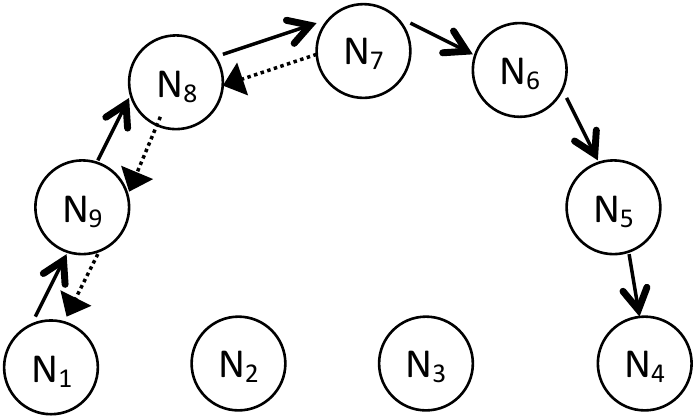}
  \caption{Joint OR and IXNC.}
  \label{fig:notBeneficialJoint}
\end{subfigure}
\caption{Traditional routing outperforms the joint approach.}
\label{fig:notBeneficial}
\end{figure}

It is clear that IXNC improves the network performance by exploiting coding opportunities as much as possible. However, when integrated with OR, maximizing coding opportunities in each transmission does not necessarily guarantee the best performance. The example provided in \figurename~\ref{fig:notBeneficial} explains the reasoning behind this statement. In this figure, nodes $N_1$ and $N_7$ transmit their packets to $N_4$ and $N_1$, respectively, while for each node only the nodes next to it are in its transmission range.   

As shown in \figurename~\ref{fig:notBeneficial}(a), the shortest path for these two flows are totally separated without any coding opportunity. On the other hand, in the joint approach, node $N_1$ selects $N_2$ and $N_9$ in its forwarder set, while forwarding via $N_9$ provides more coding opportunities than via $N_2$. Obviously, if the joint approach attempts to blindly maximize coding opportunities, it would wrongly choose $N_9$ as the next forwarder, and make the packets originated from $N_1$ travel a too long path to reach the destination, as shown in \figurename~\ref{fig:notBeneficial}(b). Therefore, for the joint approach to be effective, it should avoid straying packets far away from the shortest path.

\section{Taxonomy of Joint Protocols \label{sec:taxonomy}}
As discussed before, to develop an effective joint OR and IXNC approach some issues need to be addressed. In this section, we discuss some fundamental components of both IXNC and OR and their realization in different protocols.

\subsection{Routing metric \label{subsec:metric}}
A routing metric is used in routing protocols to determine and rank the nodes in the forwarder set. The main purpose of OR is to reduce the ExNT required to deliver packets to their final destination, leading to a shorter end-to-end delay and higher throughput~\cite{ORSurvey-Boukerche-ACMSurvey2014}. Therefore, it is very important to choose the right candidate forwarders, and prioritize them in an efficient way. This necessitates having an appropriate routing metric, which is even more critical in joint IXNC and OR approaches as not only ExNT but also the number of coding opportunities must be taken into account.

The most common metrics used in OR protocols include hop-count~\cite{ROMER-Yuan-WiMesh2005, OPRAH-Westphal-MASS2006, CAR-Liu-NetSysManage2015}, geographic distance~\cite{GeRaF-Zorzi-TMC2003, CORE-OR-Yan-IEEEWC2010, CAR-Liu-NetSysManage2015}, and link quality-based protocols like ETX~\cite{EXOR-Biswas-SIGCOMM2005, XCOR-Koutsonikolas-Mobicom2008, CAOR-Chung-ICC2012, CAOR-Yan-ICC2008, CORMEN-Islam-COMPUTING2010, O3-Han-MobiHoc2011} and EAX (Expected Anypath Count)~\cite{OAPF-Zhong-SIGMOBILE2006, Anypath-Dubois-TN2011, Anypath-Li-MASS2009, Anypath-Cerda-WoWMoM2010}. The ETX of each link is measured as the expected number of transmissions to successfully transmit a packet and receive its ACK over that link. If $P_{i,j}$ denotes the probability of successful transmission from node $i$ to node $j$, the ETX for the link between $i$ and $j$ equals 
\begin{equation}
\text{ETX} = \dfrac{1}{P_{i,j} \times P_{j,i}} \nonumber .
\end{equation}EAX, first applied in OAPF (Opportunistic Any-Path Forwarding)~\cite{OAPF-Zhong-SIGMOBILE2006}, is another routing metric that takes into account the effect of opportunistic routing in calculating the ExNT. 
In fact, for any source $s$ and destination $d$ with a given forwarder set $C^{s,d}$ selected by the source, EAX calculates the expected number of transmissions to successfully deliver a packet from $s$ to $d$ through the forwarder set. Given that the candidates in the forwarder set are ordered descendingly based on their priority, EAX for $s$ and $d$ can be calculated as
\begin{equation}
\text{EAX}(s,d)=\dfrac{1+\sum_i\text{EAX}(C_i^{s,d},d)f_i \prod_{j=1}^{i-1}(1-f_j)}{1-\prod_i(1-f_i)} \nonumber ,
\end{equation}
where $C_i^{s,d}$ represents the $i^{th}$ candidate in the forwarder set between $s$ and $d$, and $f_i$ denotes the probability of successful transmission from $s$ to $C_i^{s,d}$.
In addition to these metrics, joint protocols need to consider the coding gain coming from the IXNC component, either as the number of flows that are mixed or the number of neighbors that can decode the packet.

In summary, routing metrics that take into account the link quality, such as ETX and EAX, shown to improve the performance in comparison to the metrics based on hop-count or geographical distance. In terms of the coding gain, on one hand, forwarding coded packets with more coding partners or more flows combined can improve the performance due to the benefits of ``free-riding''. On the other hand, forwarding coded packets with larger forwarder set (i.e., more neighbors that can decode the packet) can improve the performance because of the path diversity provided by OR. However, as discussed in Section~\ref{subsec:shrinkage}, the forwarder set might shrink when the number of coding partners increases. Hence, a perfect routing metric not only should take into account the specifications of both OR and IXNC, but also needs to consider this trade-off as well as all other challenges discussed in Section~\ref{sec:jointDrawbacks}.

\subsection{Forwarder set coordination \label{subsec:coordination}}
By selecting more than one potential forwarder, OR provides more chances for a packet to progress toward the destination. However, in each transmission, only one of the nodes in the forwarder set (i.e., the node with the highest priority that has received the packet) should forward the packet, and other nodes should discard it. Otherwise, there would be many duplicate packets in the network degrading its performance. In fact, one needs to ensure that the extra coding opportunities in the joint approach are created because of exploring more paths and not from the duplicate packets~\cite{XCOR-Koutsonikolas-Mobicom2008}. Therefore, it is crucial to have an effective method to coordinate the nodes in the forwarder set such that they can agree on the next forwarder among themselves and avoid duplicate transmissions. 

To deal with duplicate packets, most joint IXNC and OR protocols apply a strict scheduling to coordinate forwarders. Each node sets a forwarding timer according to its priority in the forwarder set, and transmits the packet at timer expiration unless it receives a signal from a higher priority node indicating the transmission of the packet. The signaling solutions are either control-based or data-based~\cite{ORSurvey-Boukerche-ACMSurvey2014}. In data-based methods~\cite{BEND-Zhang-CNJournal2010, CORE-OR-Yan-IEEEWC2010, CAR-Liu-NetSysManage2015, FlexONC-Kafaie-TVT2017}, the nodes in the forwarder set cancel their transmission after overhearing the transmission of the same data packet by a higher-priority candidate, while in control-based approach~\cite{CORMEN-Islam-COMPUTING2010, INCOR:Inter+opp-Zhu-ICC2015} the higher priority node sends a control packet (e.g., ACK or probe) to notify others about receiving the packet. 

However, the use of a scheduler increases the delay and also requires some modification at the MAC-Layer, which causes the loss of some desirable features of IEEE 802.11. Furthermore, if the channel condition is not perfect or some of the nodes in the forwarder set are not in the transmission range of each other, they cannot overhear packet transmissions by others. This makes coordination by scheduling even further complicated.
Hence, in some studies, IANC is incorporated with OR, along with IXNC~\cite{CAOR-Chung-ICC2012, O3-Han-MobiHoc2011} or without it~\cite{MORE-Chachulski-SIGCOMM2007, CodeOR-Lin-ICNP2008, CCACK-Koutsonikolas-INFOCOM2010, ONCR-Xiang-INFOCOM2015, PipelineOR-Lin-VTC2010}, to eliminate the need of a scheduler for per-packet coordination between nodes in the forwarder set. IANC adds randomness to OR and omits the possibility of useless duplicate packets by making all packets equally beneficial to the destination through RLNC. However, note that IANC by itself, without integration with IXNC, cannot improve the network throughput beyond the capabilities of OR~\cite{Inter+opp-Mehmood-ComputNetw2013}. Further details on the comparison of these techniques are provided in Section~\ref{subsec:comparison}.

\subsection{Forwarder set selection \label{subsec:FSselection}}
\begin{figure}
\centering
\begin{subfigure}{.35\textwidth}
  \centering
  \includegraphics[width=0.98\linewidth]{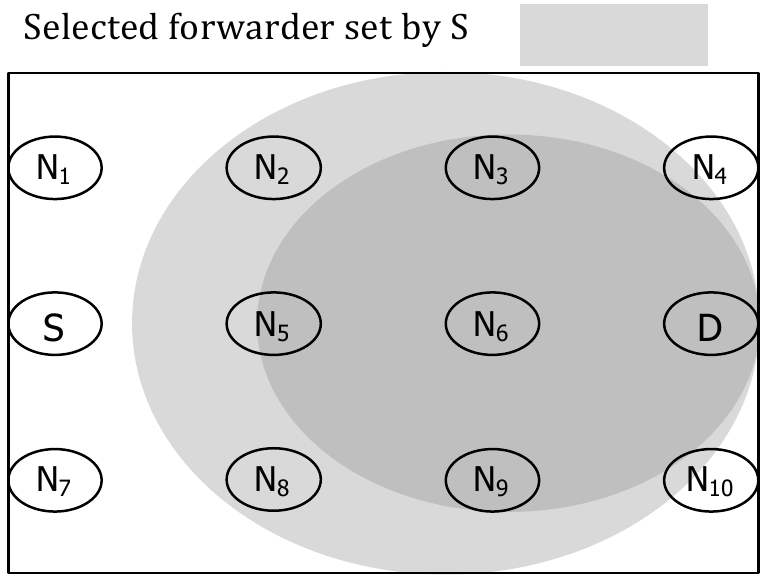}
  \caption{End-to-end approach.}
  \label{fig:endToEnd}
\end{subfigure}%
\begin{subfigure}{.35\textwidth}
  \centering
  \includegraphics[width=0.98\linewidth]{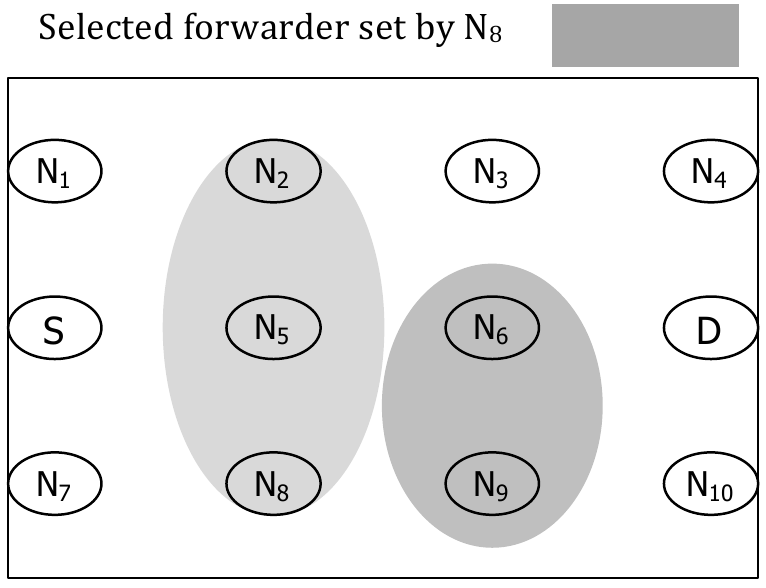}
  \caption{Hope-by-hop approach.}
  \label{fig:hopByHop}
\end{subfigure}
\caption{End-to-end versus hop-by-hop forwarder set selection.}
\label{fig:FSS_approach}
\end{figure}

Each packet usually carries the information of the forwarder set. Regarding the size of the forwarder set, while a larger one means better chance of packet progress toward the destination, it also usually means a greater overhead of coordination as well as more duplicate packets in the network~\cite{CORMEN-Islam-COMPUTING2010}. The selection of the nodes in the forwarder set can be either end-to-end or hop-by-hop\cite{NCSurvey-Iqbal-NCAJournal2011, ORSurvey-Liu-CM2009}. In end-to-end forwarder set selection, the set of potential forwarders is determined by the source once for the whole path toward the destination. On the other hand, in hop-by-hop forwarder set selection, each node determines the forwarder set toward the destination independently. For example, let us assume that in \figurename~\ref{fig:FSS_approach} $S$ transmits packets to $D$, while each node can receive packets only from the nodes immediately next to it horizontally, vertically, or diagonally. In the end-to-end approach, $S$ picks the nodes closer to $D$ than itself (based on a routing metric) and attaches this list to the packet header as the forwarder set (e.g., $\{D, N_{10}, N_4, N_6, N_9, N_3, N_5, N_8, N_2 \}$). Then, if say $N_8$ forwards the packet, it updates the forwarder set by removing those nodes which are not closer to $D$ than itself, and the new forwarder set might equal $\{ D, N_{10}, N_4, N_6, N_9, N_3, N_5 \}$. On the other hand, in the hop-by-hop approach, $S$ creates the forwarder set from the nodes closer to $D$ in its neighborhood like $N_2, N_5$ and $N_8$. Then, if $N_8$ were to forward the packet, it will choose the forwarder set independently based on the nodes in its own neighborhood like $N_6$ and $N_9$.

While an end-to-end approach covers a broader area and provides more chances for a packet to progress, its overhead is higher and its implementation is harder than a hop-by-hop strategy. Also, the coordination among forwarders is more difficult in an end-to-end strategy, and can cause duplicate transmissions as some nodes may not overhear each other. However, the end-to-end approach usually outperforms the hop-by-hop approach capitalizing on more network-wide state information~\cite{ORSurvey-Liu-CM2009}.

Note that in the joint IXNC and OR approach, when a coded packet is forwarded, the forwarder sets of coding partners (i.e., decoding forwarder sets) must be disjoint sets; otherwise, those packets cannot be coded together~\cite{HCOR-Hai-EURASIP2014}. That is, for each coding partner $p$, the nodes in its forwarder set are its potential next-hops that have already overheard all other coding partners and can decode $p$. If any of these nodes, say $N$, is in the forwarder set of another coding partner $\bar{p}$, it contradicts the fact that $N$ has already overheard $\bar{p}$. Thus, these nodes cannot be in the forwarder set of any other coding partner.

\begin{figure}[ht]
\centering
\includegraphics[scale=0.9]{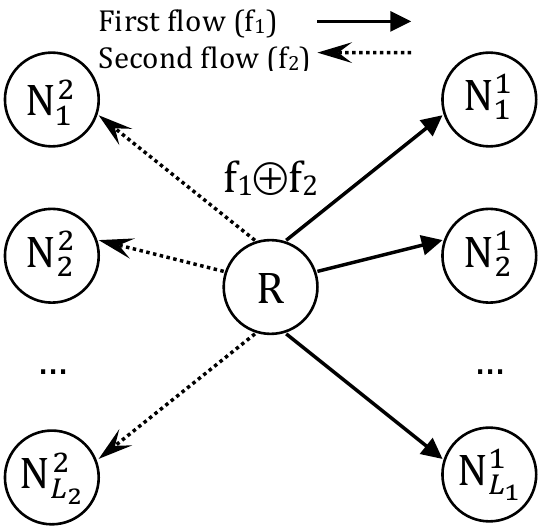}
\caption{The decoding forwarder sets must be disjoint sets.}
\label{fig:disjointFS}
\end{figure}

To illustrate the idea, let us assume that in \figurename~\ref{fig:disjointFS} to forward a coded packet $f_1 \oplus f_2$, $R$ selects $C^1_R = \lbrace N^1_1, N^1_2, ..., N^1_{L_1} \rbrace$ and $C^2_R = \lbrace N^2_1, N^2_2, ..., N^2_{L_2} \rbrace$ as the decoding forwarder sets of $f_1$ and $f_2$ respectively, where $L_i$ denotes the size of the decoding forwarder set of $f_i$. Let us assume that these two sets are not disjoint; then there exist $N^1_k \in C^1_R $ and $N^2_t \in C^2_R $ such that $N^1_k = N^2_t$. Let us call this node, which is in the decoding forwarder set of both coding partners, $N$. If $N \in C^1_R$ then it has not received $f_1$ so far; otherwise it should have forwarded $f_1$ instead of $R$. Similarly, $N$ could not have received $f_2$. Now, if $N$ has heard neither $f_1$ nor $f_2$, how is it supposed to decode $f_1 \oplus f_2$? Therefore, $N \notin C^1_R  \text{ AND } N \notin C^2_R$, which shows these decoding forwarder sets must be disjoint.

In summary, most of the existing joint IXNC and OR protocols adopt the hop-by-hop forwarder set selection strategy as it requires less complex selection algorithms and coordination methods. In addition, in contrast to the end-to-end strategy, the hop-by-hop strategy does not need global topology information and works based on the neighborhood information, which is already available in all protocols. Regarding disjoint decoding forwarder sets, coding conditions, discussed in Section~\ref{subsec:codingCond}, ensure that the combined packets are decodable, and their decoding forwarder sets are disjoint.

\subsection{Coding region \label{subsec:region}}

The majority of research on IXNC is limited to a two-hop region. This means a node will encode a packet with other packets if the next hop of the packet is able to decode it. In fact, if the next hop cannot decode the packet, it will drop the packet. However, in some topologies, even if the next hop can not decode the packet, it could still forward it as coded, and another node further down stream might be able to perform decoding successfully. 

\begin{figure}[ht]
\centering
\includegraphics[scale=0.9]{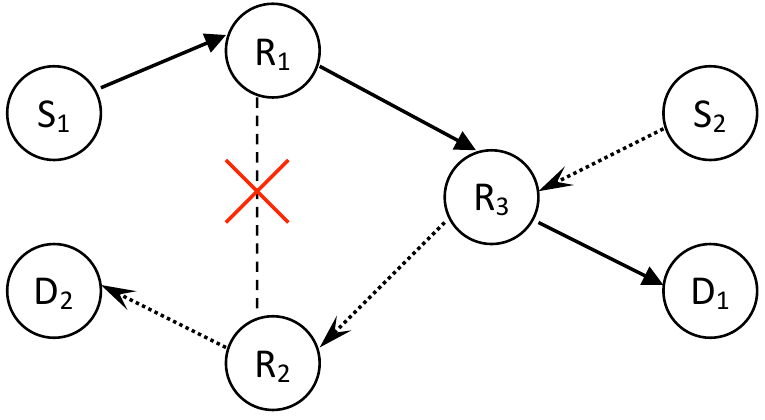}
\caption{Coding opportunities beyond a two-hop region.}
\label{fig:codingRegion}
\end{figure}

We use the scenario depicted in \figurename~\ref{fig:codingRegion} to demonstrate the idea.
In this figure and in a 3-hop network, $S_1$ and $S_2$ send their packets to $D_1$ and $D_2$, respectively. Although the destinations can overhear the packets of the other flow from its source, in a two-hop region IXNC cannot find any coding opportunity in such a topology. Note that $R_{2}$, which is the next hop of the packets of the second flow (i.e., flow from $S_2$ to $D_2$), cannot overhear any packet from the first flow. Thus, if $R_{3}$ encodes packets of these two flows, $R_{2}$ cannot decode the packet. However, in multi-hop coding, $R_2$ forwards the received coded packet, and then $D_2$ will decode it using the packets overheard from $S_1$. In fact, if nodes have access to information about the network topology and the route of the flows, IXNC protocols will be able to capture such multi-hop coding opportunities and benefit from them.

The implementation of multi-hop coding seems more challenging in the joint approach than in the regular IXNC. Of course, for a coding node to be aware of the nodes that have already received a packet, the list of the upstream nodes of the packet can be embedded in the packet header. However, predicting downstream nodes is more complicated as it does not include only the nodes on the shortest path. In fact, in joint OR and IXNC approach even the next-hop is not specified, let alone the other downstream nodes. Hai \textit{et al.}~\cite{HCOR-Hai-EURASIP2014} attempt to implement multi-hop coding in the joint approach, but the only node, in addition to the neighbors of the coding node (i.e., potential next-hops), which is considered in the coding decisions is the final destination.

\subsection{Coding conditions \label{subsec:codingCond}}
In IXNC methods, the performance of the network depends on the decodability of the coding opportunities found in the network. To ensure the coded packet $a \oplus b$ is decodable, in coding within a two-hop region, the next hop of each packet should have already received the other packet. In a more general coding region (i.e., multi-hop coding), there should be a node in downstream of each packet that has already received the other packet. 

Usually reception reports are sent by every node to inform others about the packets that have already been received by the node. The reception reports are piggybacked on data packets or broadcast periodically as control packets. Each node makes decisions on encoding based on the information provided by the reception reports of other nodes. However, there are cases that nodes cannot rely on such deterministic information for encoding because of the loss of reception reports or their late arrival. In such cases, probabilistic information is used~\cite{COPE-Katti-IEEEACMTransactions2008}. 

\begin{figure}[ht]
\centering
\includegraphics[scale=0.9]{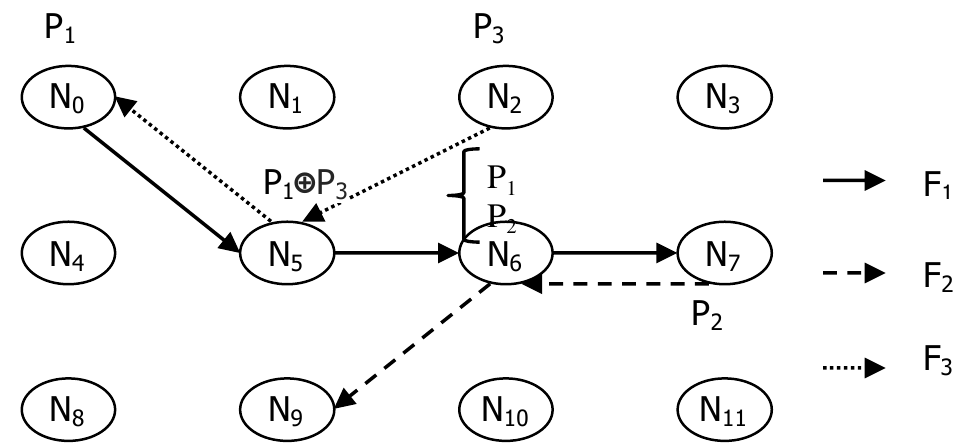}
\caption{Problem with the coding conditions~\cite{FlexONC-Kafaie-TVT2017}.}
\label{fig:codingCondProb}
\end{figure}

For coding in a two-hop region, this means when the link quality between nodes is greater than a threshold, two packets are combined if the next-hop of each packet is the previous-hop of the other packet or one of the neighbors of the previous-hop. For example, in \figurename~\ref{fig:Xtopology}, $R$ can combine $a$ and $b$ because the next-hop of $a$ (i.e., $D_1$) is a neighbor of the previous-hop of $b$ (i.e., $S_2$), and also the next-hop of $b$ (i.e., $D_2$) is a neighbor of the previous-hop of $a$ (i.e., $S_1$).
For mixing more than two flows, every two of them should hold the conditions above. However, in some scenarios as shown in~\cite{FlexONC-Kafaie-TVT2017}, encoding decisions made based on these coding conditions may decide incorrectly to mix some packets that cannot be decoded at the next-hops.

To elaborate on the inaccuracy of these coding conditions, let us assume that in \figurename~\ref{fig:codingCondProb}, reproduced from ~\cite{FlexONC-Kafaie-TVT2017}, $N_0$ transmits $P_1$ to $N_7$, $N_7$ transmits $P_2$ to $N_9$, and $N_2$ transmits $P_3$ to $N_0$. Also, we assume that these sources are out of transmission range of their destinations, and $N_5$ forwards $P_1$ and $P_3$ as a coded packet $P_1 \oplus P_3$. 
Then, $N_6$ decodes $P_1$, and based on these coding conditions, the combination of $P_1$ and $P_2$ at $N_6$ seems a valid encoding strategy.
% because the next-hop of $P_1$ (i.e., $N_7$) is the previous hop of $P_2$, and the next-hop of $P_2$ (i.e., $N_9$) is one of the neighbors of the previous hop of $P_1$ (i.e., $N_5$). 
However, if $N_6$ forwards the coded packet $P_1 \oplus P_2$, $N_9$ cannot decode $P_2$ correctly as it has only overheard  $P_1 \oplus P_3$ but neither $P_1$ nor $P_3$. As explained in~\cite{FlexONC-Kafaie-TVT2017}, the problem happens because the previous hop of $P_1$ (i.e., $N_5$) sends it as a coded packet; therefore, its neighbors (e.g., $N_9$) do not receive $P_1$ natively.

Adding OR to IXNC makes coding conditions even more complicated. From one perspective, since there is not just one single next-hop for each packet, the chance that packets can be combined is higher. In fact, as long as at least one of the nodes in the forwarder set of each packet has already overheard the other packet, they can be encoded together. From another perspective, the forwarder set of the coding partners (i.e., decoding forwarder set) might be smaller than their original forwarder set, as explained in Section~\ref{subsec:shrinkage}, because it should only include those nodes that can decode the packet.

\subsection{Coding strategy \label{subsec:codingStrategy}}
In IXNC methods, all native packets are usually stored at the same output queue. However, for the sake of easier access, some research separates packets in different virtual queues based on their corresponding flow or their next-hop (i.e., a virtual queue is assigned to the packets of each flow or the packets that are to be forwarded to the same nex-hop)~\cite{COPE-Katti-IEEEACMTransactions2008, XCOR-Koutsonikolas-Mobicom2008}.
In addition, coded packets are either generated when a transmission opportunity is available (i.e., on-demand) or beforehand (i.e., prepared). 

\begin{figure}[ht]
\centering
\includegraphics[scale=0.8]{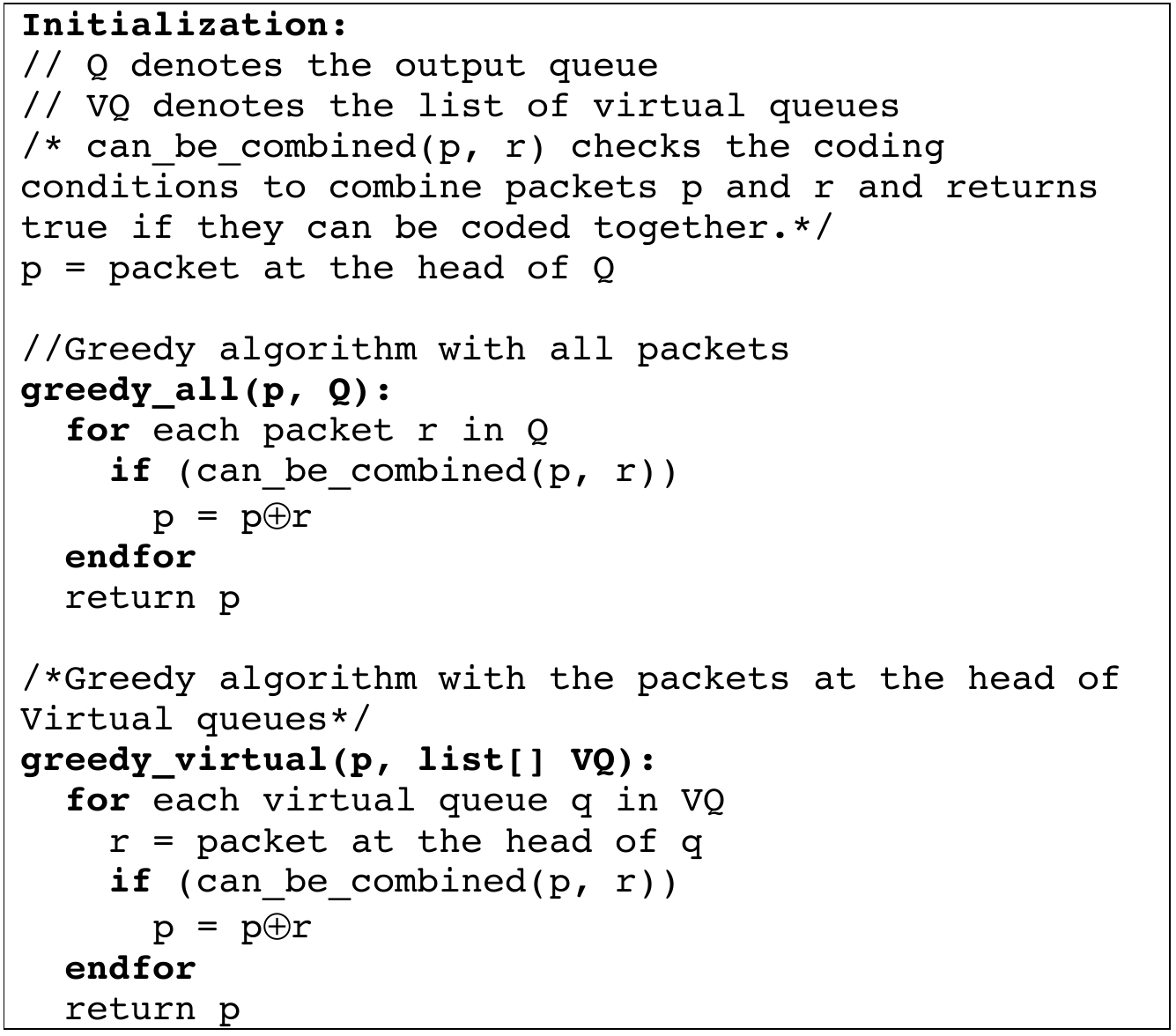}
\caption{Greedy algorithms for finding coding partners.}
\label{fig:encodingGreedy-pseudo}
\end{figure}

\begin{figure}[ht]
\centering
\includegraphics[scale=0.8]{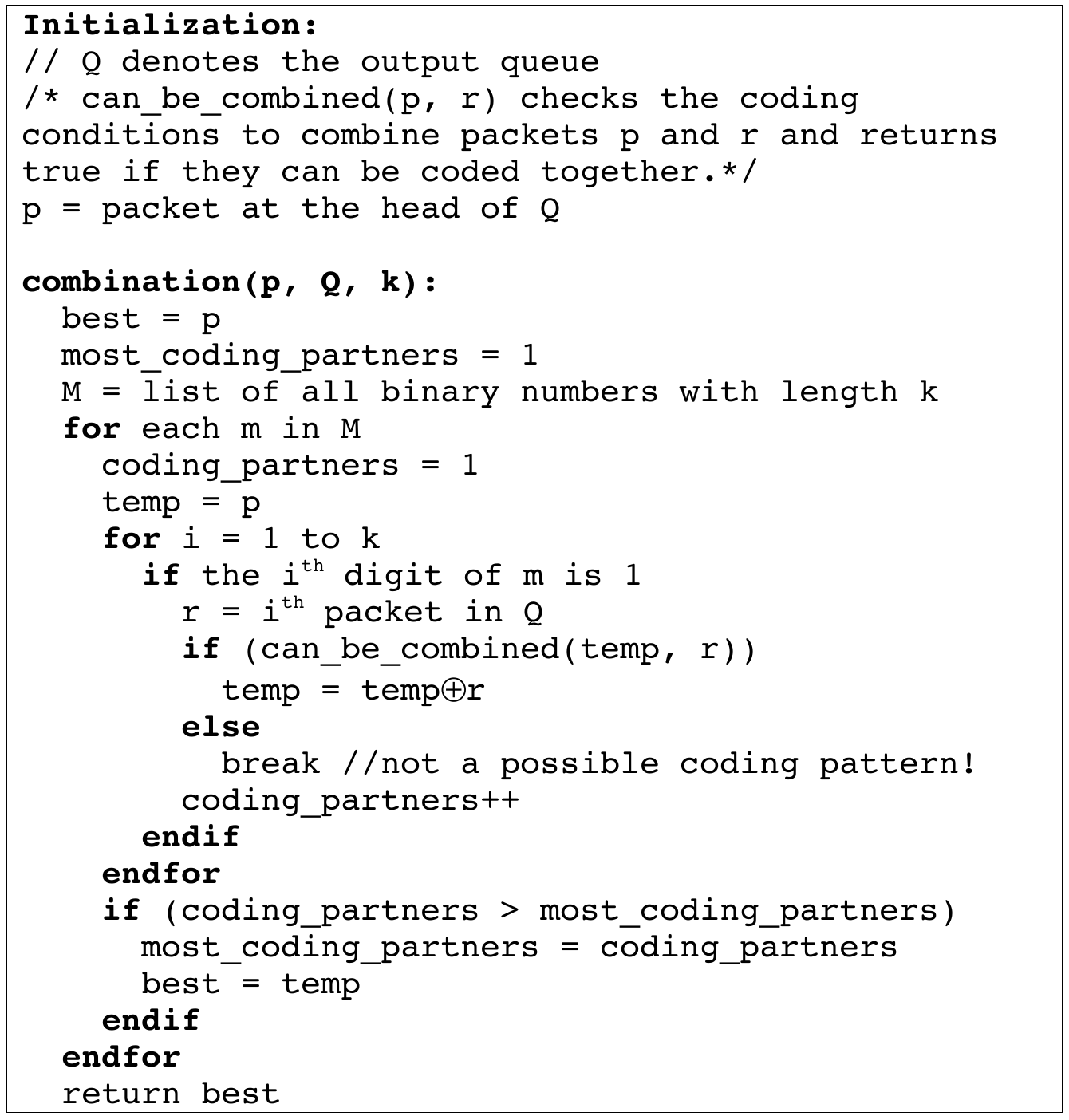}
\caption{Exploring the first $k$ packets of the output queue to find the largest number of coding partners for the packet at the head of output queue.}
\label{fig:encodingComb-pseudo}
\end{figure}

In an on-demand approach, all packets are stored as native packets in the output queue, and when there is a transmission opportunity, the node chooses the native packet at the head of the queue (i.e., called first packet), encodes and transmits it. 
To maximize the number of coding partners combined with the first packet, it is required to explore all possible combinations of the first packet with the packets of the output queue. Since, it is computationally expensive, some studies apply heuristic methods or search only the first $k$ packets of the output queue for all potential encoding patterns.
Also, some studies consider a greedy algorithm by sequentially searching either all the packets of the output queue or only the packets at the head of virtual queues.  \figurename~\ref{fig:encodingGreedy-pseudo} and \figurename~\ref{fig:encodingComb-pseudo} present the pseudo-code for the mentioned search algorithms.

To avoid delaying a transmission for searching queues and finding coding opportunities, in a prepared approach, as soon as a new packet arrives at the output queue, the node mixes it with other packets if there is any coding opportunity. Hence, an output queue is assigned to the coded packets; coded packets might also be assigned to separate (virtual) queues based on the number of their coding partners. To find the potential coding partners for the new arrived packet, usually first the coded queue and then the native queue are searched. In fact, IXNC aims to generate coded packets with maximum coding partners because the larger the number of coding partners, the higher the ``free-riding'' and coding gain. Note that similar search algorithms as in the on-demand approach can be applied in the prepared approach. In addition, when a node has a transmission opportunity, it usually gives the higher priority to the packet at the head of the coded queue.

The on-demand approach may capture more coding opportunities (i.e., coded packets with more coding partners) as the coding decisions are based on the latest information before transmission. However, it imposes more encoding delay, which can be considerably high especially when all possible combinations of the packets are explored. Therefore, in delay-sensitive applications, the prepared approach seems more feasible. On the other hand, the on-demand approach might be used in cases where the delay is not the most important factor, or coding partners are found using either greedy algorithms or the all-combinations algorithms with small $k$.

\subsection{ACK strategy \label{subsec:ack}}
Since the next-hop is not predetermined in OR, packets are usually broadcast, which eliminates the link layer implementation of the acknowledgment. Therefore, additional mechanisms are required, either at link or network layer, to recover lost packets and provide reliable transmissions. Two main available cases for packet recovery are hop-by-hop ACKs and end-to-end ACKs. End-to-end ACKs are generated by the final destination and impose less overhead but increase the end-to-end delay. On the other hand, hop-by-hop ACKs are generated by each forwarder and reduce the delay at the cost of higher overhead~\cite{diversitySurvey-Bruno-CompComm2010}.

Furthermore, coded packets in IXNC are forwarded to more than one next-hop, and if all these next-hops were to acknowledge their corresponding packet immediately after receiving it from the sender, there would be a high probability of collision among them. To avoid this issue, COPE uses pseudo-broadcast (i.e., emulated broadcast) transmissions in which the link layer destination of the packet header is set to the address of one of the next-hops, and additional fields are added to the packet header for the next-hops of other coding partners. Then hop-by-hop ACKs are sent asynchronously by all the next hops, added to the header of ready-to-be-sent data packets or in periodic control packets. Therefore, to reduce the feedback overhead, a group of packets might be acknowledged together via bitmaps or reception reports piggybacked on data packets or sent periodically as control packets. 

\subsection{Opportunistic coding \label{subsec:opCoding}}
Opportunistic coding in IXNC means a packet is sent natively if there is not any coding opportunity. In this case, coded packets are generated opportunistically and without postponing transmission of native packets artificially to generate coded packets. However, in some IXNC studies, the transmission of the native packet, ready to be forwarded, is delayed to receive its coding partner(s), and send coded packets instead of native ones as much as possible. This approach provides more coding opportunities, and simplifies estimating the rate of coding opportunities at each node~\cite{AnaNC-Kafaie-TMC2017}.
However, not applying opportunistic coding causes a longer end-to-end delay, especially with the asymmetric flows as the transmission of native packets should be held, waiting for coding partners to arrive. 
In general, it seems more reasonable to apply opportunistic coding in the joint approach. First it reduces the delay. Second, in OR it is not clear beforehand which packets are received by which nodes (i.e., there is no designated next-hop). Therefore, it is not practical to postpone the transmission of native packets at a node waiting for some potential coding partners, which may never arrive at that node.

Any proposed protocol for merging IXNC and OR techniques needs to address the mentioned challenges, and make smart technical design decisions regarding these fundamental components. As discussed here, the available solutions, inherited from either IXNC or OR, can be helpful but certainly are not sufficient, and further investigation is required to consider the effect and characteristics of both techniques in realization of the joint approach.
The classification of joint IXNC and OR protocols, based on these main components, is summarized in Table~\ref{table:JointClassification}.
 
\begin{table*}
\renewcommand{\arraystretch}{1}
\caption{Taxonomy of joint OR and IXNC protocols.}
\label{table:JointClassification}
 %\begin{adjustwidth}{-1.1cm}{}
% \nointerlineskip\leavevmode
%\begin{center}
\centering
\scriptsize
%\tiny

\begin{tabular}{|c|c|c|c|c|c|c|c|c|}
\hline
\multirow{2}{*}{\textbf{Protocol}} & \textbf{Routing} & \textbf{Coordination}  & \textbf{Coding} & \textbf{Max flows}  & \textbf{Opportunistic}  & \textbf{Coding}   &  \textbf{ACK} & \textbf{Forwarder} \\
& \textbf{metric}  & \textbf{method}  & \textbf{region} & \textbf{mixed} & \textbf{coding} & \textbf{strategy}  & \textbf{strategy} & \textbf{set selection} 
 \tabularnewline
\hline
\hline
\multirow{5}{*}{XCOR~\cite{XCOR-Koutsonikolas-Mobicom2008}} & \multirow{5}{*}{ETX}  &  & \multirow{5}{*}{Two-hop}  & \multirow{5}{*}{Multi} & \multirow{5}{*}{Yes} & On-demand, &  & \multirow{5}{*}{End-to-end} \\
 &   &  Timer, &  &  &  & max utility gain & Reception &  \\
 &   &  data-based &  &  &  & by checking flows & reports &  \\
 &   &  &  &  &  & in desc.\ order &  & \\
  &   &  &  &  &  & of Q's length & & \\
\hline
\multirow{3}{*}{CAOR~\cite{CAOR-Yan-ICC2008}} & ETX, & Timer,  & \multirow{3}{*}{Two-hop} & \multirow{3}{*}{Multi}  & \multirow{3}{*}{-}  &  On-demand, & Reception &  \multirow{3}{*}{Hop-by-hop} \\ 
 & coding &  data-based &   &  &  & comb.\ of first & reports  &  \\ 
  & partners &   &   &  &  & $k$ packets &  &\\ 
\hline
\multirow{3}{*}{ANCHOR~\cite{ANCHOR-Jiao-WiCom2008}} & No. of & Notification, & \multirow{3}{*}{Two-hop} & \multirow{3}{*}{Multi}  & \multirow{3}{*}{No} & Prepared, & \multirow{3}{*}{-} & \multirow{3}{*}{-} \\ 
 & transmissions &  data-based &  &   &  & greedy based &  &  \\ 
 &  &   &  &   &  & on fewest trans. &  &  \\ 
\hline
\multirow{3}{*}{CORMEN~\cite{CORMEN-Islam-COMPUTING2010}} & ETX,  & Timer,  & \multirow{3}{*}{Two-hop}  &  \multirow{3}{*}{Multi}  & \multirow{3}{*}{Yes} &  & End-to & \multirow{3}{*}{Hop-by-hop} \\ 
 & coding  & control-based  &  &   &  & Prepared  & end &  \\
 & partners  &  &  &   &  &  &  & \\
\hline
\multirow{3}{*}{CORE~\cite{CORE-OR-Yan-IEEEWC2010}} & No.\ of  & Timer,  & \multirow{3}{*}{Two-hop}  & \multirow{3}{*}{Multi} & \multirow{3}{*}{No} & On-demand,  & \multirow{3}{*}{-}  & \multirow{3}{*}{Hop-by-hop}\\ 
 & receivers,  & data-based  &   &  &  & comb.\ of first &  &  \\
  & geo.\ dist.  &   &   &  &  & $k$ packets &  & \\
\hline
\multirow{3}{*}{BEND~\cite{BEND-Zhang-CNJournal2010}} & Coding & Timer, & \multirow{3}{*}{Two-hop} & \multirow{3}{*}{Multi}  & \multirow{3}{*}{Yes} & Prepared, & Hop-by & \multirow{3}{*}{Hop-by-hop}\\ 
& partners, &  data-based &  &   &  & greedy & hop & \\ 
& ETX & & & & & & & \\
\hline
\multirow{2}{*}{O3~\cite{O3-Han-MobiHoc2011}} & \multirow{2}{*}{ETX} & \multirow{2}{*}{IANC} &  \multirow{2}{*}{Two-hop} & \multirow{2}{*}{Two} & \multirow{2}{*}{No} & On-demand, & End-to & \multirow{2}{*}{End-to-end} \\ 
 &  &  &   &  &  & greedy & end &  \\ 
\hline
\multirow{3}{*}{AONC~\cite{AONC-Shen-DSN2012}} & Maximum & \multirow{3}{*}{Timer}  & \multirow{3}{*}{Tow-hop} & \multirow{3}{*}{Multi}  & \multirow{3}{*}{Yes} & Prepared, & Hop-by & \multirow{3}{*}{Hop-by-hop} \\ 
 & space &   &  &   &  & maximizing & hop &  \\ 
 & utilization &   &  &   &  & space utilization &  &  \\  
\hline
CAOR~\cite{CAOR-Chung-ICC2012} & ETX & IANC & Two-hop & Multi  & -  & -  & - & - \\ 
\hline
\multirow{3}{*}{CoAOR~\cite{CoAOR-Hu-GLOBECOM2013}} & ETX, & Timer,  & \multirow{3}{*}{Tow-hop} & \multirow{3}{*}{Multi}  & \multirow{3}{*}{Yes} & On-demand, & Reception & \multirow{3}{*}{Hop-by-hop} \\ 
 & coding &  data-based &  &   &  & comb.\ of first & reports &   \\ 
 & partners &  &  &   &  & $k$ packets &  & \\ 
\hline
\multirow{3}{*}{\cite{Inter+opp-Mehmood-ComputNetw2013}} & Probability of  & \multirow{3}{*}{-$^*$} & \multirow{3}{*}{Two-hop} & \multirow{3}{*}{Two} & \multirow{3}{*}{Yes} & \multirow{3}{*}{-} & \multirow{3}{*}{Hop-by-hop} & \multirow{3}{*}{Hop-by-hop} \\
& successful &  &  & & & & & \\
& transmission &   &  & & & & & \\
\hline
\multirow{3}{*}{HCOR~\cite{HCOR-Hai-EURASIP2014}} & Anypath  & \multirow{3}{*}{-}  & Two-hop & \multirow{3}{*}{Two}  & \multirow{3}{*}{Yes} & Prepared,  & \multirow{3}{*}{-}  & Hop-by-hop\\ 
 & cost  &  & +   &   &  & greedy &  &  +\\ 
  &   &  & destination &   &  &  &  & destination\ \\ 
\hline
\multirow{3}{*}{INCOR~\cite{INCOR:Inter+opp-Zhu-ICC2015}} & \multirow{3}{*}{CETX} & Timer, & \multirow{3}{*}{Two-hop} & \multirow{3}{*}{Two}  & \multirow{3}{*}{Yes} & On-demand, & Hop-by & \multirow{3}{*}{Hop-by-hop} \\ 
 &  &  control-based &  &   &  & checking first & hop &  \\ 
  &   &   &   &  &  & $k$ packets &  & \\
\hline
\multirow{4}{*}{CAR~\cite{CAR-Liu-NetSysManage2015}} & coding &   &  \multirow{4}{*}{Two-hop} &  \multirow{4}{*}{Multi}  &  \multirow{4}{*}{No} & Prepared, &  \multirow{4}{*}{-} & \multirow{4}{*}{Hop-by-hop} \\ 
 & partners,  &  Timer, &  &   &  & greedy &  &   \\ 
  & geo.\ dist.,  & data-based &  &   &  & grouping flows &  & \\ 
  & hop-count  &  &  &   &  &  &  & \\ 
\hline
\multirow{4}{*}{FlexONC~\cite{FlexONC-Kafaie-TVT2017}} & ETX, & Timer & \multirow{3}{*}{Two-hop} & \multirow{3}{*}{Multi}  & \multirow{3}{*}{Yes} & Prepared, & Hop-by & \multirow{3}{*}{Hop-by-hop} \\
& coding & data-based & & & & greedy & hop & \\
& partners & &  & & &  & & \\
\hline
\end{tabular}
%\end{center}
%* It is assumed that the nodes have enough information to agree on the same forwarder. 
% \end{adjustwidth}
\end{table*}

\section{Review of Proposed Joint Protocols \label{sec:review}}
The possibility of combining OR and IXNC was first discussed in~\cite{ImportanceofOR-Katti-2005}, where a preliminary version of COPE was introduced as well. However, the results suggested that the benefit of combining these two techniques is not notable, and even duplicate packets can degrade the network performance in some scenarios. In that early research, forwarders are prioritized based on their distance from the destination, and coding opportunities are not taken into account. Also, the coordination among forwarders is not discussed in details.

\subsection{Coordination using IANC}
CAOR (Coding-Aware Opportunistic Routing)~\cite{CAOR-Chung-ICC2012} is one of those few studies that utilize IANC as the coordination method of OR in realizing the joint approach. In each transmission, CAOR combines the packets of flows that maximize a metric, which is defined in terms of the progress of the packet in each transmission (based on ETX) and the probability that the next-hops will receive the coded packet and decode it. However, the throughput gain of CAOR is relatively smaller than that of the other joint methods~\cite{CAR-Liu-NetSysManage2015} mainly because combining IANC and IXNC reduces the number of coding opportunities in the network.

O3 (Optimized Overlay-based Opportunistic routing)~\cite{O3-Han-MobiHoc2011} is another approach that exploits IANC in integration of OR and IXNC, where packets of not more than two flows can be mixed. In O3, an overlay network performs overlay routing, IANC and IXNC, while in underlay network OR is applied, and an optimization problem is solved to find the desirable sending rates for IANC and IXNC packets. Using Qualnet simulation, the results show that O3 outperforms shortest path routing, COPE and MORE. Note that while in regular IANC, only the final destination needs to decode RLNC packets, joint approaches discussed here (i.e., CAOR and O3) impose more overhead and longer delay as all intermediate nodes need to apply Gaussian elimination and decode RLNC packets (to decode IXNC packets).

\subsection{Coordination using timer}
\subsubsection{End-to-end forwarder set selection}
One of the first studies on joint OR and IXNC is XCOR (Interflow NC with Opportunistic Routing)~\cite{XCOR-Koutsonikolas-Mobicom2008}, whose OR component has been inspired by SOAR (Simple Opportunistic Adaptive Routing)~\cite{SOAR-Rozner-2006}. In XCOR, the forwarder set, which forms a ``thin belt'' along the shortest path, is calculated recursively for each next-hop by the source and stored in the packet header. Also, the forwarders are prioritized based on their closeness to the destination in terms of ETX. Before forwarding a received packet, the forwarders start a timer according to their priority, and cancel the packet transmission if they overhear it from a higher-priority node. Also, cumulative reception reports, in the form of a bitmap, provide feedback for local recovery and packet mixing.

To find the best coded packet at each node, XCOR defines a utility function as the sum of the utility gain of the next-hops, which is calculated in terms of the progress toward the destination, the probability of successful transmission to the next-hop, and the probability of successful decoding at the next-hops. Applying a heuristic algorithm, they rank flows in terms of the length of their output queues, and mix the packet at the head of the longest one with the packet at the head of other flows' queues if this combination increases the utility gain. In the evaluation of XCOR in Qualnet, two simple topologies (i.e., a hexagon topology and a chain topology with 4 nodes) are considered, and results show that its performance degrades significantly in lightly loaded or lossy environments~\cite{CAOR-Chung-ICC2012}.

\subsubsection{End-to-end ACK}
In CORMEN (Coding-aware Opportunistic Routing in wireless Mesh Network)~\cite{CORMEN-Islam-COMPUTING2010}, as an IXNC scheme enhanced with OR, the nodes in the forwarder set should have a good quality link with the sender (in terms of ETX), and the ETX of the link between any pair of them is within a threshold. Also, to avoid diverging the path and unnecessary duplicate packets, the nodes in the forwarder set are neighbors of the nodes in the shortest path. In CORMEN, end-to-end ACKs are sent instead of hop-by-hop ones, and each forwarder starts a forwarding timer in terms of ETX and the maximum number of flows that can be mixed in a coded packet. When the timer reaches zero, the node will notify others by sending a probe packet, and other nodes will cancel their timer and transmission of the corresponding packet. Similar to source routing protocols, the packet header should contain not only the forwarder set but also the nodes on the shortest path. In addition, since the packet may not follow the shortest path, the forwarders need to keep updating the path. 

\subsubsection{Geo-position as the routing metric}
CAR (Coding-Aware opportunistic Routing)~\cite{CAR-Liu-NetSysManage2015} is another joint scheme that aims to maximize the number of native packets coded together in a single transmission by dynamically selecting the route based on real-time coding opportunities. Regarding encoding, CAR keeps a set of coding groups representing the flows that can be potentially coded together. In CAR, each node knows the geographic position of all other nodes in the network, and the nodes in the forwarder set are neighbors of the sender that 1) their hop-count to the destination is less than or equal to the sender, and 2) are closer to the destination than the sender (in terms of the geographic distance). Each node sets the forwarding timer inversely proportional to the number of native packets in a coded packet, and nodes cancel their transmission after overhearing the same packet from another node in the forwarder set. Also, native packets are only sent by the next-hop designated by the shortest path routing. For TCP (i.e., Transmission Control Protocol) flows, ACK packets are sent along the shortest path and are coded only with themselves.

In another work, considering geographic distance as the routing metric, CORE (Coding-aware Opportunistic Routing)~\cite{CORE-OR-Yan-IEEEWC2010} selects the forwarder set from the neighbors of the sender which are geographically closer to the destination than itself.
The main components of CORE are forwarder set selection, coding opportunity calculation, primary forwarder selection (i.e., calculating local coding opportunities by each node), and priority-based forwarding (i.e., using timers by nodes to coordinate among themselves). In each transmission, among all nodes in the forwarder set, CORE selects the node with the highest coding gain as the next forwarder. To prioritize the nodes with different coding opportunities, forwarding timers are used so that the node with more coding opportunities forwards its packet earlier. In addition, in CORE each packet carries the location of the sender and the destination, and no retransmission mechanism for lost packets has been described in their work. To forward a packet at the head of the output queue, CORE picks the next $k$ packets as seeds for possible encoding, and chooses the one that maximizes the coding gain.

\subsubsection{Considering link quality in coding gain }
While CORE defines the coding gain function at each node in terms of the number of candidates in the forwarder set that are able to decode a coded packet, CoAOR (Coding-Aware Opportunisitc Routing)~\cite{CoAOR-Hu-GLOBECOM2013} takes into account the number of flows coded in a packet, the link quality and the number of nodes that are able to encode and decode packets as well. Analytic Hierarchy Process (AHP)~\cite{AHP-Saaty-MngSci1986} is applied to find the weight of these parameters. Three main components of CoAOR are coding-aware forwarder set selection, node coding gain calculation, and priority-based packet forwarding. The candidates in the forwarder set are selected from the neighbors of the sender closer to the destination than the sender itself (in terms of ETX), which are able to overhear each other. They coordinate among themselves using a forwarding timer inversely proportional to their coding gain.

\subsubsection{Diffusion gain}

\begin{figure}[ht]
\centering
\includegraphics[scale=0.8]{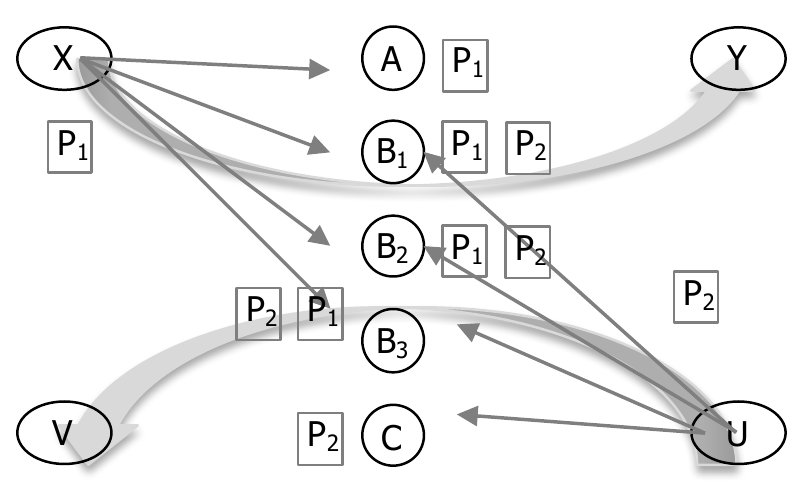}
\caption{Diffusion gain in BEND~\cite{BEND-Zhang-CNJournal2010}.}
\label{fig:diffusion}
\end{figure}

BEND~\cite{BEND-Zhang-CNJournal2010}, as another joint approach, introduces a type of gain, referred to as the \emph{diffusion gain}, which is the benefit of being able to scatter flows through multiple forwarders dynamically.
To avoid traffic concentration in BEND, the neighbors of the sender closer to the destination may receive a native packet and mix and forward it on behalf of the next-hop designated by the routing protocol, where DSDV (i.e., Destination-Sequenced Distance-Vector)~\cite{DSDV-Perkins-1994} is used as the routing protocol.
For example in \figurename~\ref{fig:diffusion}, where nodes $A$ and $C$ are the next-hops of the flows from $X$ to $Y$ and from $U$ to $V$, respectively, BEND allows $B_1$, $B_2$ and $B_3$, which can overhear packets of both flows to combine and forward the packets on behalf of $A$ and $C$. To do so, a \emph{second-next-hop} field is included in the header of native packets. As such, when a neighbor of the sender receives a native packet, it can find the address of the next-hop in the \emph{second-next-hop} field. 
However in BEND, OR cannot be applied to two consecutive hops, and coded packets can only be received and decoded by the designated next-hops on the shortest path.  

FlexONC (Flexible Opportunistic Network Coding)~\cite{FlexONC-Kafaie-ICC2015, FlexONC-Kafaie-TVT2017} is another joint IXNC and OR approach that considers a union of the packets of the neighborhood to create coding opportunities, while packets travel around the shortest path. In FlexONC, the nodes in the forwarder set of a packet are the neighbor of both the next-hop and the second next-hop of the packet.
By maintaining forwarding tables of all neighbors, instead of carrying \emph{second-next-hop} information by each packet, FlexONC is able to address the described issues of BEND.
Moreover, this research discovers that the conditions used in previous studies to combine packets of different flows are overly optimistic and would affect the network performance adversarially. Therefore, a more accurate set of rules is provided for packet encoding. Using simulations in NS-2, FlexONC is compared with traditional non-coding scheme, COPE, CORE and BEND. The experimental results show that FlexONC outperforms other baselines especially in networks with high bit error rate, by better utilizing redundant packets permeating the network, and benefiting from precise coding conditions.

\subsubsection{Application in wireless multimedia sensor networks (WMSN)}
In a study focusing on WMSNs, AONC (Adaptive Opportunistic Network Coding)~\cite{AONC-Shen-DSN2012} improves the transmission quality of video streams. 
In regular IXNC, if coding partners do not have the same size, the shorter packet is padded with zeros. Therefore, the length of the coded packet is determined by the length of the longer coding partner. This reduces the coding space utilization, which is a metric representing the amount of original data carried by a coded packet of a certain length.
Given that video packets have variable lengths, to resolve this issue and encode coding partners with similar lengths, AONC might send more than one packet of a flow in each transmission. In fact, to maximize the forwarded length, it splices packets of the same flow as long as the spliced packet's length is less than the space length limit. Then, the spliced packets of different flows could be mixed using IXNC. Their optimization algorithm is repeated for different coding groups and different possible space length limits, and finally the one with maximum space utilization is selected. Although this method reduces the number of required transmissions, it intensifies the packet reordering problem of OR.

\subsubsection{Coding-based ETX}
INCOR (Inter-flow Network Coding-based Opportunistic Routing)~\cite{INCOR:Inter+opp-Zhu-ICC2015}
introduces a metric called Coding-based Expected Transmission Count (CETX) that computes the expected transmission count required to deliver one packet to a destination using IXNC. This metric is used to prioritize the nodes in the forwarder set when lower CETX means higher priority. To calculate CETX for all nodes, they run a centralized algorithm similar to Dijkstra's algorithm based on the idea that adding the closer neighbors of node $i$ and their forwarder set (in terms of CETX) to the forwarder set of node $i$ can reduce its CETX. Each packet carries its prioritized forwarder set, and the forwarders start a forwarder timer proportional to their priority, which will be canceled upon hearing an ACK from a higher-priority node.

\subsubsection{Coding is not always the best option}
HCOR (High-throughput Coding-aware Opportunistic Routing)~\cite{HCOR-Hai-EURASIP2014} is a distributed system based on anypath routing~\cite{anypath-Laufer-IATN2012} claiming that maximizing coding opportunities does not necessarily improve the network performance. 
Since the forwarder set of a coded packet is a subset of the original forwarder set (i.e., the subset that can decode the coded packet), HCOR argues that sending coded packets is not always beneficial, and one may need to decide if IXNC decreases the cost by ``free-riding'' or increases it by shrinking forwarder set. Therefore, they consider coding gain as well as link qualities to find the path with minimal anypath cost. In HCOR, at most two packets can be mixed, and the coordination among nodes in the forwarder set has not been discussed.
They compare HCOR with anypath routing as well as a joint OR and IXNC approach that always encodes packets if there is any coding opportunity, referred to as COOR in their research. The results show that HCOR outperforms other baselines in different scenarios by $10\%$ to $30\%$.

\subsection{Coordination using reception reports}
A few studies on the joint approach provide information for the nodes by exchanging reception reports and control packets such that ideally all nodes act consistently having a similar picture of the network, and can agree on the same best forwarder without using any timer or applying IANC. This coordination technique can cause inconsistency of information among nodes leading to a sub-optimal forwarder selection, deadlock (i.e., no one transmits the packet), or non-decodable transmissions. To handle these issues, techniques like guessing (i.e., using the delivery probability of the links as the probability that the neighbor has received a packet) or timers are used as well. 
 
ANCHOR (Active Network Coding High-throughput Optimizing Routing)~\cite{ANCHOR-Jiao-WiCom2008} is a method in which packets carry the shortest path information. By exploiting coding opportunities, ANCHOR actively updates the route, which has been embedded in the packet header. Based on reception reports, if a node other than the next-hop of the packet can provide more coding opportunities, it notifies the other nodes to update the route. Simulation results in Glomosim show that ANCHOR performs better than COPE and DSR~\cite{DSR-Johnson-2007} in a number of scenarios.

In another method called CAOR~\cite{CAOR-Yan-ICC2008}, the nodes in the forwarder set are neighbors of the sender closer to the destination than the sender (in terms of ETX) that can mutually overhear each other. To find the higher-priority forwarder with most coding opportunities, nodes exchange reception reports advertising not only their own stored packets but also their neighbors' packets. Doing so, all nodes in the forwarder set can compute available coding opportunities in each other and will know their transmission priorities, and which one of them is the best forwarder for this particular transmission. To avoid duplicate packets, nodes will cancel their transmission if they overhear another node in the forwarder set transmitting the same packet. Also, to compensate for lost or delayed reception reports, each node guesses about packets it would receive; if a node has received $M$ consecutive packets of a flow, it can report the next two packets of that flow in its current reception report. Nevertheless, CAOR still uses timers, and each node sets its transmission timer proportional to its priority. Simulation results of comparing CAOR and COPE in a $1000 \times 1000$ field with $200$ nodes shows that as the number of UDP (i.e., User Datagram Protocol) flows in the network increases CAOR outperforms COPE due to providing more coding opportunities. However, when the number of flows is considerably large, they perform similarly because there is little room for CAOR to increase coding opportunities by forwarding packets through a path other than shortest one. Also, by increasing the number of hops between the source and destination, CAOR's throughput drops not as quickly as that of COPE because of providing more coding opportunities and alleviating the effect of congestion.

In a more theoretical study, Mehmood \textit{et al.}~\cite{Inter+opp-Mehmood-ComputNetw2013} discuss the optimal approach in a combination of IXNC and OR for bidirectional unicast flows between two nodes relayed by multiple common neighbors. They argue that naive combination of OR and NC is sometimes sub-optimal, and may even perform worse than NC over a shortest path. They propose a dynamic programming algorithm to find a lower bound on the expected number of transmissions required to communicate a packet in both directions in terms of link error probabilities. In their model, all nodes are in interference range of each other and concurrent transmissions are not possible in the network. Also, all nodes report the information state (i.e., the packets overheard or received) of themselves and their neighbors in a series of ACKs after transmission of every data packet (ACK cycle). They assume that the source nodes are always saturated, at any given time only one packet of each flow exists in the network, and all nodes already know the link probability parameters in advance. 

To find the best scheduling plan to deliver one packet of both sources to their destinations, the network state in each time instance is calculated, which includes the information state of all nodes representing the packets overheard by them. Their model provides a DAG (directed acyclic graph), where each nodes denotes a network state and any edge corresponds to a transmission of a packet (from one of two flows or their combination), and a subset of the nodes receiving that transmission. For each network state, a score is defined as the expected number of transmissions from that state until both destinations receive their packets. An algorithm similar to Bellman-Ford algorithm for shortest path (or Viterbi algorithm for finding a maximum-likelihood sequence of states) is applied to find an edge (i.e., transmission) with which the score of the new node is the minimum possible (i.e., referred to as optimal action). By applying this algorithm recursively, eventually it finds the optimal score for the starting state, where two sources are to transmit their packets. The complexity of the algorithm is $(3K+2)4K+1$, where $K$ is the number of intermediate nodes. Using numerical results (without considering the effect of ACK cycle overhead), optimal scheme can reduce the number of required transmissions by up to $20\%$ comparing with OR or IXNC in isolation, or their simple combination.

\subsection{Summary and comparison of proposed joint protocols}\label{subsec:comparison}
As discussed in this section, the most common coordination method is using timers to schedule the transmission at the candidate forwarders. The candidates set their forwarding timer in terms of their rank, which can be pre-determined and attached to the packet header by the sender or calculated by each node based on the available local information (e.g., the proximity to the destination, the coding gain of the scheduled transmission). However, since this scheduling is not usually strict, the candidates also rely on random medium access, and each node cancels its transmission if it overhears the transmission of the same packet by another node in the forwarder set. Therefore, it is crucial that all nodes in the forwarder set can overhear each other over high-quality links.

To avoid the issues related to the timers, a forwarder set can be coordinated using IANC in which, instead of scheduling the candidates to forward only the packets that have not been transmitted by higher priority nodes, the candidates forward random combination of their received packets. Doing so, all packets are equally beneficial, and the destination can decode the packets after receiving enough innovative packets. However, this increases the end-to-end delay due to the encoding and decoding delay. Furthermore, one of the biggest challenges of IANC is deciding on the number of coded packets that each candidate should forward. While sending unnecessary packets degrades the network performance, there should be sufficient transmissions to decode the original native packets at the destination~\cite{diversitySurvey-Bruno-CompComm2010, ORSurvey-Boukerche-ACMSurvey2014}. To alleviate this issue and control the number of packets spread in the network, a credit-based approach is usually applied in which each node's credit represents the number of qualified transmissions, but it is not perfect. In addition, adding IXNC to IANC makes the implementation even more challenging as each coded packet can be both RLNC of the packets of the same flow and \textit{XOR}ed of the packets of different flows. The design of protocols that implement these two layers of encoding and decoding is tricky, especially by considering the fact that IXNC packets usually are decoded at the next-hop, while the decoding of the IANC packets are done only by the destination. 

Furthermore, as described in this section, joint protocols propose various metrics to decide on the nodes in the forwarder set and prioritize them. However, they often agree on the incorporation of the closeness to the destination (in terms of ETX, hop-count or geo-distance) and the coding gain (usually defined as the number of combined flows or the number of neighbors that can decode the packet). While these two parameters should be taken into account, some other criteria are required for the joint approach to outperform the individual IXNC and OR approaches. For example, any joint protocol needs to ensure that after each transmission, the packet becomes closer to the destination. On ther other hand, choosing a perfect metric, which selects the best candidate at each transmission, requires more state information to be gathered at each node and increases the overhead and the complexity of the protocol especially if more than two flows are combined. Therefore, further studies are needed to develop an efficient joint approach based on the characteristics of these two techniques as well as some other factors such as the network topology and the traffic pattern~\cite{ORSurvey-Liu-CM2009}. Table~\ref{table:JointComparison} summarizes the comparison of mentioned joint protocols with highlights of their performance including their principal idea, the simulation tool used, their evaluation scenarios and their throughput gain.

\begin{table*}
\caption{Comparison of joint OR and IXNC protocols.}
\label{table:JointComparison}
 \begin{center}
 %\scriptsize
 \tiny
\begin{tabular}{|c|c|c|c|c|c|c|}

\hline
\multirow{2}{*}{\textbf{Protocol}}  & \textbf{Principal} & \textbf{Simulation} &	\textbf{Evaluation}  & \textbf{TCP} & \textbf{Compared}  & \textbf{Throughput} \\ 
& \textbf{idea} & \textbf{tool}  & \textbf{topology} & \textbf{/UDP} & \textbf{protocols} & \textbf{gain}  \\ 
\hline
\hline
\multirow{5}{*}{XCOR~\cite{XCOR-Koutsonikolas-Mobicom2008}} & One of the first & \multirow{5}{*}{Qualnet} &  & \multirow{5}{*}{UDP} &  & By $115\%$, $34\%$ and $13\%$  \\ 
 & methods showing  &   & 4-node chain, &  & SOAR~\cite{SOAR-Rozner-2006}, &  in chain, and $75\%$, $22\%$  \\ 
  & benefits of the &   & hexagon &  & COPE,  &  and $70\%$ in hexagon,  \\ 
 & joint approach &  &  &  & Srcr~\cite{Srcr-Bicket-MobiCom2005} &  it outperforms Srcr, \\ 
  &  &   &  &  & &  SOAR and COPE \\ 

\hline
\multirow{4}{*}{CAOR~\cite{CAOR-Yan-ICC2008}} & Coding gain by & \multirow{4}{*}{NS-2}  &   & \multirow{4}{*}{UDP} & \multirow{4}{*}{COPE} &   \\ 
 & forwarding based on &   & $200$ nodes in  &  &  & On average $15\%$  \\ 
 & coding opportunity  &   & $1000 \times 1000$ $m^2$  &  &  & improvement  \\
 & awareness  &   &   &  &  &   \\
\hline
\multirow{3}{*}{ANCHOR~\cite{ANCHOR-Jiao-WiCom2008}} & Optimizing route & \multirow{3}{*}{Glomosim}  & $100-700$ nodes  & \multirow{3}{*}{UDP} & \multirow{3}{*}{COPE, DSR~\cite{DSR-Johnson-2007}}  & outperforms COPE    \\ 
 & based on no.\ &   & in $800 \times 800$ $m^2$  &  &   &  by up to $38\%$  \\ 
  & of transmissions  &   &   &  &   &    \\ 
\hline
\multirow{3}{*}{CORMEN~\cite{CORMEN-Islam-COMPUTING2010}} &  Finding the & \multirow{3}{*}{NS-2}  & $3 \times 3$, $5 \times 3$ and  & \multirow{3}{*}{UDP} & \multirow{3}{*}{COPE}  & Outperforms COPE   \\ 
 & shortest coding  &   & $5 \times 5$ grids  &  &   &  slightly  \\ 
  & possible path  &   &   &  &   &    \\ 
\hline
%CORE~\cite{CORE-OR-Yan-IEEEWC2010} & ???  & NS-2  & ???  & COPE, OR\protect\footnotemark
\multirow{3}{*}{CORE~\cite{CORE-OR-Yan-IEEEWC2010}} & Forwarding based on & \multirow{3}{*}{NS-2}  & $200$ nodes in  & \multirow{3}{*}{UDP} & \multirow{3}{*}{COPE, OR} & On average $22\%$  \\ 
 & maximizing coding &   & $800 \times 800$ $m^2$  &  &  & improvement  \\ 
 & opportunities  &   &   &  &  &   \\ 
\hline
\multirow{3}{*}{BEND~\cite{BEND-Zhang-CNJournal2010}} & Neighborhood coding & \multirow{3}{*}{NS-2}  & Cross, 3-tier  & \multirow{3}{*}{UDP} & COPE, & On average about $25\%$  \\ 
 & repository around &   & $5 \times 5$ grid  &  & traditional & improvement over COPE  \\ 
 & the shortest route &   &   &  & routing &    \\ 
\hline
\multirow{5}{*}{O3~\cite{O3-Han-MobiHoc2011}} & Sending rate & \multirow{5}{*}{Qualnet}  & 3-node chain, diamond,   & \multirow{5}{*}{UDP} &   &   \\ 
&  optimization in &   & $5 \times 5$ grid,  &  & COPE, MORE,  & Significantly   \\ 
& joint IANC, OR &   & 25-node random,  &  & traditional & outperforms   \\ 
&  and IXNC &   & MIT Roofnet~\cite{roofnet},  &  & routing  &  other baselines \\ 
&  &   & UW testbed~\cite{UW-Reis-SIGCOMM2006}  &  &   &   \\ 

\hline
\multirow{3}{*}{AONC~\cite{AONC-Shen-DSN2012}} & Better video & \multirow{3}{*}{NS-2}  & Cross, 3-tier,  & \multirow{3}{*}{UDP} & COPE, BEND, & Higher quality  \\ 
 & transmission &  & $5 \times 5$ grid  &  & traditional & video trans.  \\
  & quality in WMSNs &  &    &  & routing  &   \\
\hline
\multirow{3}{*}{CAOR~\cite{CAOR-Chung-ICC2012}} & Joint OR, IANC and & \multirow{3}{*}{-}  & 2-tier with 8 &  \multirow{3}{*}{UDP} & \multirow{3}{*}{MORE} & $20\% - 27\%$  \\ 
 & IXNC to increase &  & nodes, random &  &  & improvement  \\ 
  & coding opportunities &  &  &  &  &   \\ 
\hline
\multirow{4}{*}{CoAOR~\cite{CoAOR-Hu-GLOBECOM2013}} & Coding gain based & \multirow{4}{*}{MATLAB}  &   & \multirow{4}{*}{UDP} & \multirow{4}{*}{CAOR~\cite{CAOR-Yan-ICC2008}} &  \\ 
 & on link quality, coding &   & 200 nodes in  &  &  & $2\% - 30\%$  \\ 
  & partners and  &   & $1000 \times 1000$ $m^2$  &  &  & improvement \\ 
   & decoding ability  &   &  &  &  &  \\ 
\hline
\multirow{4}{*}{HCOR~\cite{HCOR-Hai-EURASIP2014}} & Deciding & \multirow{4}{*}{NS-2}  & 3-node chain, cross, & \multirow{4}{*}{UDP} & anypath routing,  &  \\ 
 & between coding &  & hexagon, diamond,  &  & COOR (HCOR & $10\% - 30\%$ \\ 
 & and native &  & 50 nodes in  &  & without calculating   & improvement \\ 
 & transmissions &  & $1000 \times 1000$ $m^2$ &  & IXNC cost)  &  \\ 
\hline
\multirow{4}{*}{INCOR~\cite{INCOR:Inter+opp-Zhu-ICC2015}} & Forwarder set  & \multirow{4}{*}{-} &  & \multirow{4}{*}{UDP} & \multirow{4}{*}{COPE, EAX~\cite{EAX-Zhong-2007}} & Outperforms COPE and\\ 
 & selection and  &  & $5 \times 5$ grid,  &  &   & EAX , on average  \\ 
  & prioritization  &  & MIT Roofnet~\cite{roofnet} &  &   & by $12\%$ and $17\%$  \\ 
  & based on CETX  &  &  &  &   &   \\ 
\hline
\multirow{4}{*}{CAR~\cite{CAR-Liu-NetSysManage2015}} & Maximizing the & \multirow{4}{*}{NS-2}  &  & \multirow{4}{*}{Both} & \multirow{4}{*}{COPE, BEND}  & Cross, TCP: $43\%$(COPE),  \\ 
 &  no. of native &   &  Cross,  &  &   &  $36\%$(BEND). Cross, UDP:   \\ 
  & packets coded in &   & $5 \times 5$ grid  &  &   &  $34\%$(COPE), $15\%$(BEND)  \\
   & each transmission &   &   &  &   &    \\ 
\hline
\multirow{4}{*}{FlexONC~\cite{FlexONC-Kafaie-TVT2017}} & NC around shortest & \multirow{4}{*}{NS-2} & 2-tire and 3-tire & \multirow{4}{*}{UDP} & COPE, BEND & On average, by $19\%$, $94\%$, \\
& path with precise & & multi-hop, & & CORE, & $29\%$ in 2-tire and $23\%$, $32\%$,\\
& encoding conditions & & $5 \times 5$ grid & & traditional routing &  $41\%$ in grid, it outperforms \\
& & & & & & BEND, CORE, COPE \\
\hline
\end{tabular}
\end{center}
\end{table*}

\vspace{0.5 cm}

\section{Issues, Challenges, and Future Research Directions \label{sec:future}}
 As discussed in Section~\ref{sec:review}, some methods have been proposed to combine IXNC and OR; however, most of them are yet to fully utilize the broadcast nature of wireless networks. In some described works, the closeness to the destination (i.e., to find the forwarder set) is calculated in terms of the hop count or the geographic distance~\cite{CAR-Liu-NetSysManage2015, CORE-OR-Yan-IEEEWC2010}, which does not necessarily represent the quality of the path. In addition, in many studies, the path traveled by the node can be excessively longer than the shortest path~\cite{CAOR-Yan-ICC2008, CORE-OR-Yan-IEEEWC2010, CoAOR-Hu-GLOBECOM2013, CAR-Liu-NetSysManage2015, INCOR:Inter+opp-Zhu-ICC2015}, which can increase the end-to-end delay, and degrade the performance. Even most of those studies that take into account the length of the route and select the forwarder set from nodes around the shortest path cannot combine packets of more than two flows~\cite{HCOR-Hai-EURASIP2014} or require the source to know the shortest path and embed it in the packet header~\cite{XCOR-Koutsonikolas-Mobicom2008, CORMEN-Islam-COMPUTING2010, ANCHOR-Jiao-WiCom2008}. 

Furthermore, as discussed in Section~\ref{sec:jointDrawbacks}, although integration of IXNC and OR seems promising, a simple and perfunctory combination of these two components does not necessarily outperform each individually, and even may degrade the network performance. As we explained in Section~\ref{sec:taxonomy}, any IXNC or OR protocol must make important design decisions, and address some important questions. However, the solutions proposed in the literature for the realization of each of these two techniques, in isolation, may not fit their combination. Thus, to have a synergistic effect in this integration, one may need to revisit the implementation and objectives of each component accordingly, and consider the characteristics of both techniques in the proposed solutions for the challenges described at the end of Section~\ref{sec:motivation}.

Therefore, further research on the idea of integrating IXNC and OR is imperative to not only better capture the coding opportunities in the network, but also control effectively how far packets stray away from a designated shortest path.
To this end, in addition to exploring the key components discussed in Section~\ref{sec:taxonomy}, some other research directions on further improving the performance of joint IXNC and OR approach can be briefly outlined as follows:

\begin{itemize}

\item \emph{Adding IANC} -- By applying IANC, the joint approach can omit the need to a strict scheduler for forwarder set coordination, as explained in Section~\ref{subsec:coordination}. In recent years, a number of publications have augmented the joint approach with IANC~\cite{CAOR-Chung-ICC2012, O3-Han-MobiHoc2011}; 
however, further studies are required to efficiently merge these three great techniques, and address some important challenges about the way and the complexity of merging two separate encoding techniques (i.e., IANC and IXNC) at different nodes 

\item \emph{Coding beyond a two-hop region} -- There has been some research on extending the coding region in the network by proposing a new NC-aware routing protocol~\cite{DCAR-Le-ICDCS2008, GCCpostpone-Guo-ICC2010, FORMpostpone-Guo-VT2011}, but not in conjunction with OR. To the best of our knowledge, in the only study on the joint approach with multi-hop coding, coded packets cannot be decoded at any other node than the next-hop or the final destination~\cite{HCOR-Hai-EURASIP2014}. Thus, it would be of interest to strengthen IXNC by including a combination of OR and more powerful detection of coding opportunities beyond a two-hop region.

 \item \emph{Working properly with TCP} -- In general, NC significantly supports UDP flows, but for TCP flows, it may achieve a gain much lower than expected because of the congestion control mechanism in TCP windows. However, in recent years a few studies have been conducted to control sent and received packets and ACKs to the transport layer, so that NC can be applied without much effect on TCP windows~\cite{TCPNC-Sundararajan-IEEE2011, TCPNC-Hassayoun-INFOCOM2010, TCPNCOR-Zhang-PeerJCS2016, TCPFNC-Sun-ICC2015, ComboCoding-Chen-AdvResearch2011, TCPVON-Bao-GlobeCom2012}. Hence, a future extension of the joint approach could be its exploration and modification under TCP flows.
 
 \item \emph{Physical-layer network coding} -- Physical-layer network coding (PNC)~\cite{PNC-Zhang-MobiCom2006, PNC-Popovski-ICC2006, PNCSurvey-Hu-BSC2010, PNC-Liew-PhyComm2013} is another type of NC, in which nodes simultaneously transmit packets to a relay node that exploits mixed wireless signals to extract a coded packet. Merging OR with PNC, instead of IXNC, can be considered as another way of combining the power of OR and NC to boost the network performance. Although many ideas from the joint IXNC and OR approach can be transferred, this integration brings its own unique challenges and benefits due to unique characteristics of PNC. For example, in PNC, nodes can transmit simultaneously to a relay node without causing collision at the cost of a larger carrier sensing range. 
 
\item \emph{Joint approach in cognitive radio networks} -- In recent years, applications of NC in cognitive radio networks (CRNs) have seen considerable attention especially because of providing more efficient and secure data transmission as well as more effective spectrum utilization~\cite{CognitiveNCSurvey-Naeem-COMST2017}. Furthermore, due to the dynamic spectrum access and channel availability in CRNs, OR performs better than traditional routing protocols. Some research~\cite{IANCORinCRN-Lin-GlobeCom2010, IANCORinCRN-Zheng-ISCIT2012, IANCORinCRN-Qin-CEE2016} also adds IANC to OR as the scheduling technique.
Regarding the joint IXNC and OR approach, a few studies~\cite{NCGeoORinCRN-Tang-Globecom2012, CROR-Zhong-GlobeCom2014} investigate the effect of their combination in CRNs and show promising improvement over traditional techniques. However, further research is required to customize the features of the joint approach and make it perfectly suitable for CRNs.
For example, mentioned routing metrics and forwarder set selection strategies might not fit well CRNs and need to be tailored based on the specifications of CRNs. In addition, further investigation is required to apply the joint approach in multi-channel CRNs with uncertain channel availability.

\item \emph{Full-duplex communications} -- Due to recent advancements in wireless technology, full-duplex (FD) communication in which a node can transmit and receive data simultaneously with the same frequency band, seems more feasible. Some studies apply FD along with IANC~\cite{IANCFD-Giacaglia-ICC2013} or PNC~\cite{PNCFD-Tabataba-TVT2012, PNCFD-Tedik-VTC2014, PNCFD-Velmurugan-WPComm2015, PNCFD-Lemos-SAM2016} in wireless networks.
Also a survey on the applications of FD in CRNs can be found in~\cite{FullDuplexCRNSurvey-Amjad-COMST2017}.
We believe incorporating FD into the joint approach can be an interesting area of future investigation especially for CRNs or when PNC is used, and it can provide more reliable and efficient communication networks.

\end{itemize}

\section{Conclusion\label{sec:conclusions}}
To improve the performance of IXNC especially under poor quality channels, its integration with OR seems promising. As discussed in Section~\ref{sec:motivation}, by applying OR not only the performance of IXNC improves in lossy networks but also it can explore more coding opportunities throughout the network. 
Furthermore, when there are multiple flows in the network, by utilizing the ``free-riding'' feature of IXNC and forwarding more than one packet in each transmission, the performance of OR can be improved.
However, the scenarios discussed in Section~\ref{sec:jointDrawbacks} show that a naive integration of IXNC and OR may not possess the synergistic effect that we expect, and may even degrade the performance in comparison to applying these techniques individually. In fact, to show the real power of the joint approach, some fundamental components of either of these two techniques, discussed in Section~\ref{sec:taxonomy}, should be designed carefully considering the effect and characteristics of both of them. To show how existing joint protocols address the challenges related to these components, we compared several joint protocols in Section~\ref{sec:review} highlighting their pros and cons. Moreover, the essential issues of the joint approach as well some important future research directions were illustrated in Section~\ref{sec:future}.

\vspace{0.45cm}

\textbf{\textit{List of acronyms and abbreviations}}
\vspace{0.15cm}

\begin{abbrv}
\small
\item[ACK]						Acknowledgment
\item[ANCHOR]				Active Network Coding High-throughput Optimizing Routing
\item[AONC]					Adaptive Opportunistic Network Coding
\item[CAOR]					Coding-Aware Opportunistic Routing
\item[CAR]						Coding-Aware opportunistic Routing
\item[CETX]						Coding-based Expected Transmission Count
\item[CoAOR]					Coding-Aware Opportunisitc Routing
\item[CORE]						Coding-aware Opportunistic Routing
\item[CORMEN]				Coding-aware Opportunistic Routing in wireless Mesh Network
\item[CRN]						Cognitive Radio Network
\item[DAG]						Directed Acyclic Graph
\item[DSDV]					Destination-Sequenced Distance-Vector
\item[EAX]						Expected Anypath Count
\item[ETX]						Expected Transmission Count
\item[ExNT]						Expected Number of Transmissions
\item[ExOR]						Extremely Opportunistic Routing
\item[FD]							Full-Duplex
\item[FlexONC]				Flexible Opportunistic Network Coding
\item[HCOR]					High-throughput Coding-aware Opportunistic Routing
\item[IANC]						Intra-flow Network Coding
\item[INCOR]					Inter-flow Network Coding-based Opportunistic Routing
\item[IXNC]						Inter-flow Network Coding
\item[MAC]						Media Access Control
\item[MORE]                  MAC-independent Opportunistic Routing and Encoding
\item[NC]						Network Coding
\item[O3]						Optimized Overlay-based Opportunistic Routing
\item[OAPF]					Opportunistic Any-Path Forwarding
\item[OR]						Opportunistic Routing
\item[PNC]						Physical-layer Network Coding
\item[RLNC]						Random Linear Network Coding
\item[TCP]						Transmission Control Protocol
\item[UDP]						User Datagram Protocol
\item[WMSN]					Wireless Multimedia Sensor Network
\item[WMN]						Wireless Mesh Network
\item[WSN]						Wireless Sensor Network
\item[XCOR]					Interflow NC with Opportunistic Routing
\end{abbrv}

%\section*{Acknowledgment}

%The authors are grateful to the anonymous reviewers and the Editor in Chief, Prof. E. Hossain, for theirconstructive comments. Furthermore, the authors would like to thank  Prof. A. Haimovich  and his research group, as well as to M. Mohammadkarimi for their help with the simulation results presented in the paper.

%\color {black}
\bibliographystyle{IEEEtran}
\bibliography{IEEEabrv,citationSurvey}
%\bibliography{citationSurvey}

\begin{IEEEbiography}[{\includegraphics[width=1in,height=1.25 in,clip,keepaspectratio]{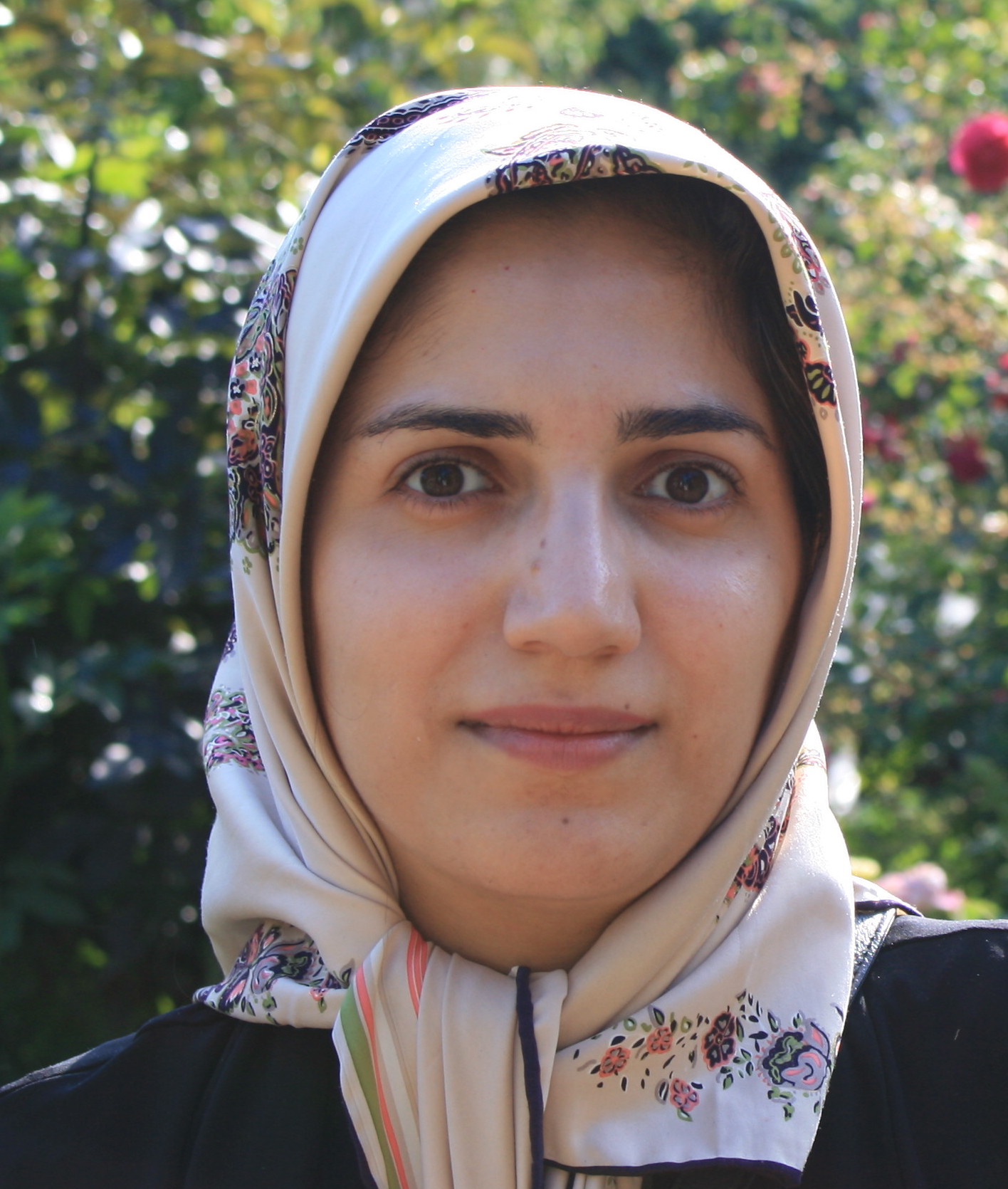}}]{Somayeh Kafaie} is currently a postdoctoral fellow in the Department of Computer Science at Memorial University of Newfoundland. She received her B.Sc. in Computer Engineering (Software) from Amirkabir University of Technology, Iran and her M.Sc. in Computer Engineering (Software) from Iran University of Science and Technology, Iran, in 2007 and 2011, respectively. She also obtained her Ph.D. in the faculty of Engineering and Applied Science, Memorial University of Newfoundland in 2017. Her research interests include wireless mesh networks, network coding, opportunistic routing, complex networks and graph theory. 
\end{IEEEbiography}

\begin{IEEEbiography}[{\includegraphics[width=1in,height=1.25 in,clip,keepaspectratio]{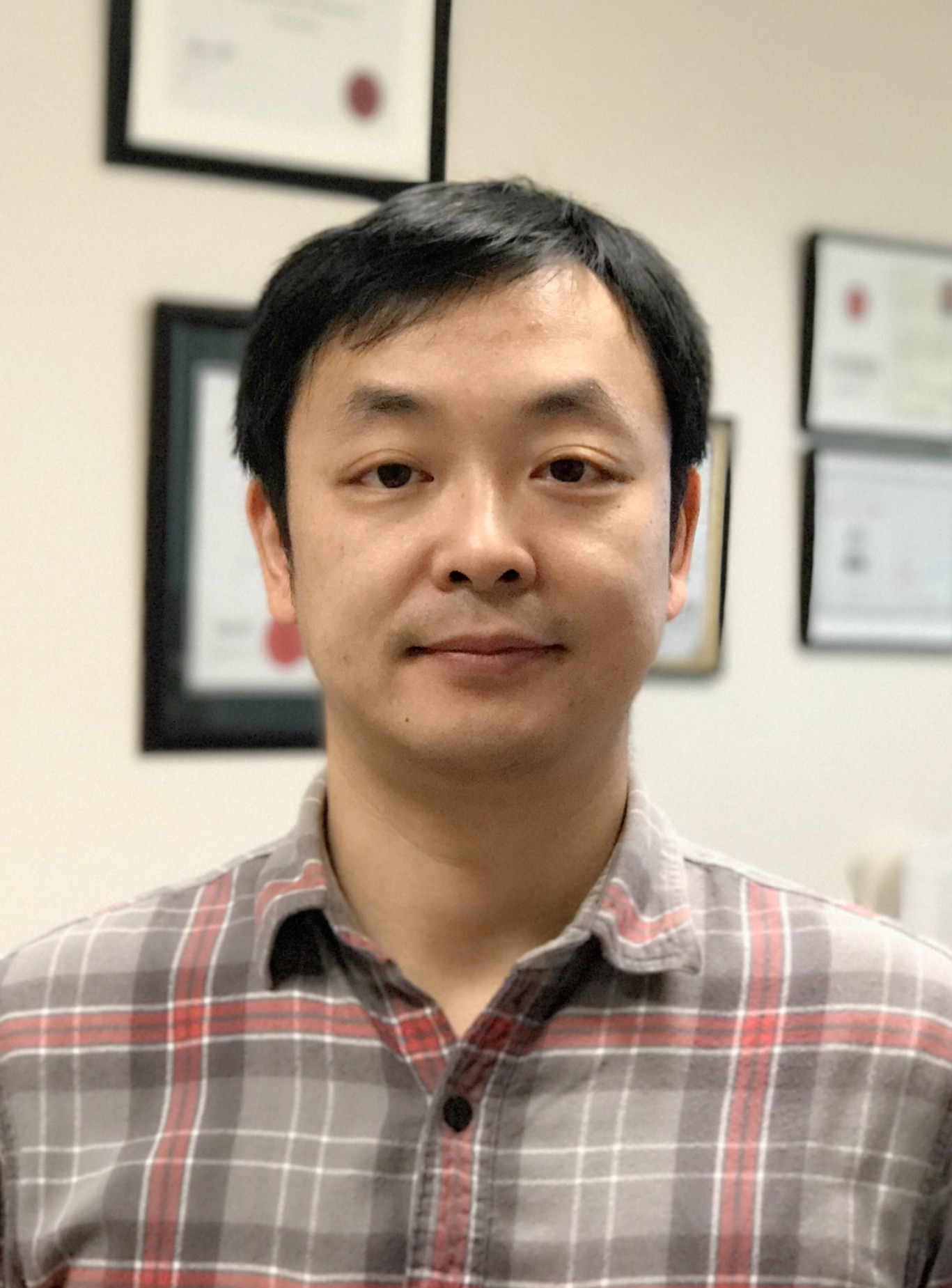}}]{Yuanzhu Chen} is an Associate Professor in the Department of Computer Science at Memorial University of Newfoundland, St. John's, Newfoundland. He was Deputy Head for Undergraduate Studies in 2012-2015, and Deputy Head for Graduate Studies in 2016 to present date.  He received his Ph.D. from Simon Fraser University in 2004 and B.Sc. from Peking University in 1999. Between 2004 and 2005, he was a post-doctoral researcher at Simon Fraser University. His research interests include computer networking, mobile computing, graph theory, complex networks, Web information retrieval, and evolutionary computation. 
\end{IEEEbiography}

\begin{IEEEbiography}[{\includegraphics[width=1in,height=1.25 in,clip,keepaspectratio]{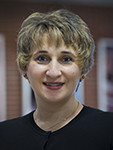}}]{Octavia A. Dobre} (M'05-SM'07) received the Engineering Diploma and Ph.D. degrees from Politehnica University of Bucharest (formerly Polytechnic Institute of Bucharest), Romania, in 1991 and 2000, respectively. She was the recipient of a Royal Society Scholarship at Westminster University, UK, in 2000, and held a Fulbright Fellowship with Stevens Institute of Technology, USA, in 2001. Between 2002 and 2005, she was with Politehnica University of Bucharest and New Jersey Institute of Technology, USA. In 2005, she joined Memorial University, Canada, where she is currently a Professor and Research Chair. She was a Visiting Professor with Université de Bretagne Occidentale, France, and Massachusetts Institute of Technology, USA, in 2013.

Her research interests include 5G enabling technologies, blind signal identification and parameter estimation techniques, cognitive radio systems, network coding, as well as optical and underwater communications among others.
Dr. Dobre serves as the Editor-in-Chief of the IEEE COMMUNICATIONS LETTERS, as well as an Editor of the IEEE SYSTEMS and IEEE COMMUNICATIONS SURVEYS AND TUTORIALS. She was an Editor and a Senior Editor of the IEEE COMMUNICATIONS LETTERS, an Editor of the IEEE TRANSACTIONS ON WIRELESS COMMUNICATIONS and a Guest Editor of other prestigious journals. She served as General Chair, Tutorial Co-Chair, and Technical Co-Chair at numerous conferences. She is the Chair of the IEEE ComSoc Signal Processing and Communications Electronics Technical Committee, as well as a Member-at-Large of the Administrative Committee of the IEEE Instrumentation and Measurement Society. Dr. Dobre is a Fellow of the Engineering Institute of Canada. 
\end{IEEEbiography}

\begin{IEEEbiography}[{\includegraphics[width=1in,height=1.25 in,clip,keepaspectratio]{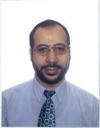}}]{Mohamed Hossam Ahmed} obtained his Ph.D. degree in Electrical Engineering in 2001 from Carleton University, Ottawa, where he worked from 2001 to 2003 as a senior research associate. In 2003, he joined the Faculty of Engineering and Applied Science, Memorial University where he works currently as a Full Professor. Dr. Ahmed published more than 135 papers in international journals and conferences. He serves as an Editor for IEEE Communication Surveys and Tutorials and as an Associate Editor for Wiley International Journal of Communication Systems and Wiley Communication and Mobile Computing (WCMC). Dr. Ahmed is a Senior Member of the IEEE. He served as a cochair of the Signal Processing Track in ISSPIT'14 and served as a cochair of the Transmission Technologies Track in VTC'10-Fall, and the multimedia and signal processing symposium in CCECE'09. Dr. Ahmed won the Ontario Graduate Scholarship for Science and Technology in 1997, the Ontario Graduate Scholarship in 1998, 1999, and 2000, and the Communication and Information Technology Ontario (CITO) graduate award in 2000. His research interests include radio resource management in wireless networks, multi-hop relaying, cooperative communication, vehicular ad-hoc networks, cognitive radio networks, and wireless sensor networks. Dr. Ahmed's research is sponsored by NSERC, CFI, QNRF, Bell/Aliant and other governmental and industrial agencies. Dr. Ahmed is a registered Professional Engineer (P.Eng.) in the province of Newfoundland, Canada. 
\end{IEEEbiography}

\end{document}